\documentclass[12pt,a4paper]{article}

\usepackage[english]{babel}
\usepackage[T1]{fontenc} 
\usepackage{lmodern}
\usepackage[cp1251]{inputenc} 
\usepackage{amsfonts,amssymb,amsmath} 
\usepackage{graphicx} 
\usepackage{cite}
\usepackage{hyperref}
\hypersetup{colorlinks=true, linkcolor=blue, filecolor=blue, urlcolor=blue} 

\allowdisplaybreaks

\makeatletter \@addtoreset{equation}{section}
\renewcommand\theequation
{\ifnum \c@section>\z@ \thesection.\fi \@arabic\c@equation}
\renewcommand{\@biblabel}[1]{#1.}
\makeatother

\newtheorem{theorem}{Theorem}    
\newtheorem{corollary}{Corollary} 

\usepackage{diagbox} 

\begin{document}

\title{\bf\Large{On optimization of coherent and incoherent controls \\ for two-level open quantum systems}}

\author{{\bf Oleg~V.~Morzhin$^{1,}$}\footnote{E-mail: {\tt morzhin.oleg@yandex.ru}; \url{http://www.mathnet.ru/eng/person30382}} \quad and \quad 
	{\bf Alexander~N.~Pechen$^{1,2,}$}\footnote{E-mail: {\tt apechen@gmail.com}; \url{http://www.mathnet.ru/eng/person17991}} \vspace{0.2cm} \\
	$^1$ Steklov Mathematical Institute of Russian Academy of Sciences,\\
	Department of Mathematical Methods for Quantum Technologies, \\
	8 Gubkina St., Moscow, 119991, Russia;\\
	$^2$ National University of Science and Technology ``MISiS'',\\
	6 Leninskiy prospekt, Moscow, 119991, Russia}

\date{}
\maketitle

\rightline{\it Dedicated to the centenary}
\rightline{\it of the birth of Academician V.S. Vladimirov}
\rightline{\it (January 9, 2023)}

\begin{abstract}
This article considers some control problems for closed and open two-level quantum systems.	The closed system's dynamics is governed by the Schr\"{o}dinger equation with coherent control. The open system's dynamics is governed by the Gorini--Kossakowski--Sudarshan--Lindblad master equation whose Hamiltonian depends on coherent control and superoperator of dissipation depends on incoherent control. For the closed system,	we consider the problem for generation of the phase shift gate
for some values of phases and final times for which numerically show that
zero coherent control, which is a stationary point of the objective functional,	
is not optimal; it gives an example of subtle point 
for practical solving problems of quantum control. 
For the open system,  in the two-stage method which was developed for generic $N$-level quantum systems in
[Pechen~A., Phys. Rev.~A., {\bf 84}, 042106 (2011)] 
for approximate generation of a target density matrix, here we consider the two-level systems for which modify the first (``incoherent'')
stage by numerically optimizing piecewise constant
incoherent control instead of using constant incoherent control analytically computed using eigenvalues of the target density matrix. Exact analytical formulas are derived for the system's state evolution, the objective functions and their gradients for the modified first stage. These formulas are applied in the two-step gradient projection method. The numerical simulations show  that the modified first stage's duration can be significantly less than the unmodified first stage's duration, but at the cost of optimization in the class of piecewise constant controls.
\end{abstract}

\section{Introduction} 

Optimal control of quantum systems (atoms, molecules, and so on) is an important
research direction at the intersection 
of mathematics, physics, chemistry, technology, which 
is crucial for development of various quantum technologies such as
quantum computing, etc.~\cite{KochEPJQuantumTechnol2022, GlaserEurPhysJD2015,
ButkovskiyBook1990, TannorBook2007, LetokhovBook2007, WizemanBook2010,BrifNewJPhys2010,MooreCS2011,FengCS2011, KochJPhysCondensMatter2016, DAlessandroBook2021}. Quantum control aims to obtain desired properties 
of the controlled quantum system using external variable influence. For example, in the theory of quantum computing, the basic objects are qubit and
quantum gate. Mathematical modeling of physical processes 
for generating quantum gates considers, e.g., 
the Schr\"{o}dinger equation with coherent control in its Hamiltonian and objective functional 
that characterizes how well a certain quantum gate is generated.
In optimal quantum control, the Pontryagin maximum principle 
(PMP)~\cite{PontryaginBook1962} (about the PMP adaptation 
for various optimal control problems of quantum systems, 
note, e.g.,~\cite[Ch.~IV]{ButkovskiyBook1990} and the survey~\cite{BoscainPRXQuantum2021}),
methods of the geometric control theory~\cite{AgrachevBook2004}, 
iterative optimization methods, including gradient-based 
methods (for quantum control, see, e.g., \cite{SchulteHerbruggenRevMathPhys2010}), 
the Krotov method (see for references the survey~\cite{MorzhinUMN2019}) are widely used. 
Control problems for two-level closed and open quantum 
systems were considered, e.g. in~\cite{AltafiniIEEE2003, GoughJOptB2005,
BonnardIEEE2009, StefanatosPRA2009, CanevaPRL2009, HegerfeldtPRL2013, ArkhincheevNonlinearDyn2017}. Various master, kinetic and transport equations are used for modelling dynamics of open classical and quantum systems. V.S. Vladimirov\footnote{Vladimirov Vasily Sergeevich (1923--2012) was a Soviet and Russian mathematician who made diverse contributions to mathematical physics, numerical methods, quantum field theory, numerical analysis, theory of generalized functions, analysis of functions of several complex variables, p-adic analysis, and multidimensional Tauberian theorems. In numerical methods, he developed the method of characteristics for solving partial differential equations (known as the Vladimirov method) and a factorization method for solving diffusion equations, proved their convergence and stability. He developed an application of the Monte Carlo method for solving problems of neutron and radiation transport and proposed a Simpson-type quadrature formula for approximate calculation of Wiener integrals.} contributed to development of numerical schemes for solving kinetic and diffusion equations, including for neutron transport~\cite{VladimirovIzvAkadNaukSSSRSerMat1957_Issue5, 
VladimirovIzvAkadNaukSSSRSerMat1958}.

An important problem is to study the dynamic landscape
of the quantum control problem, that is geometric properties 
of the corresponding objective functional
such as existence or absence of traps of various order, etc.~\cite{PechenPRA2010.82.030101,PechenEPJ2010.91.60005,PechenPRL2011,FouquieresSchirmer,PechenTrMIAN2014,PechenIzvRANSerMatem2016,PechenJPA2017,Larocca2018,Zhdanov2018,Russell2018,DalgaardPRA2021.105.012402,GeComplexSystemModelSimulation2021,VolkovJPA2021,Volkov2022}. On this topic, a rigorous proof of absence of traps was obtained for two-level coherently controlled closed quantum systems~\cite{PechenTrMIAN2014, PechenIzvRANSerMatem2016, PechenJPA2017}, which is the first and the only existing mathematically rigorous proof of trap-free behaviour for finite-level quantum systems which is also free of any assumptions. Absence of traps in the kinematic control landscape for open quantum systems was proved for the two-level case in~\cite{PechenJPA2008}, where for the first time parametrization of Kraus map open system evolution by points in the {\it complex Stiefel manifolds} was proposed and theory of optimization over complex Stiefel manifolds, including analytical computation of the gradient and Hessian of the quantum control objective, was developed for quantum control. The detailed theory for the multilevel case was developed in~\cite{OzaJPA2009}. Control landscapes for open-loop and closed-loop control were analyzed in a unified framework~\cite{PechenPRA2010.82.030101}. A~unified analysis of classical and quantum kinematic control landscapes was performed~\cite{PechenEPJ2010}. These results were used to explain unexpected, due to high dimensionality, efficiency of optimization in chemistry~\cite{MooreCS2011} and biology~\cite{FengCS2011}.

In~\cite{VolkovJPA2021}, for
the problem of generating the phase shift quantum gate, analytical
study of the objective functional's Hessian was performed at zero control, which is a stationary point of the objective functional. Also numerical exploration of the control landscape was performed using GRAPE (gradient ascent pulse engineering, see~\cite{KhanejaJMagnReson2005}) and 
stochastic methods of global optimization. The conditions of optimality of zero control and ability to exactly generate the quantum gate, including in minimal time, were found. To continue the analysis performed in~\cite{VolkovJPA2021}, in Sec.~\ref{section3} of the present work we numerically show using GRAPE, which is the most suitable to reveal structure of quantum control landscape, and also  differential evolution~\cite{StornJGlobOptim1997} and dual annealing~\cite{TsallisPhysA1996, XiangPRE2000} 
(the stochastic methods' implementations available in SciPy were used) that zero coherent control which satisfies the PMP for arbitrary amplitude constraints is not optimal; 
it thereby gives an example of subtle point for study of quantum control problems. 
In~\cite{PechenPRA2006, PechenPRA2011},  incoherent control method for quantum systems was proposed using controlled drive in the dissipator of
the Gorini--Kossakovsky--Sudarshan--Lindblad (GKSL) master equation. Such approach exploits incoherent action of the environment surrounding the system to efficiently manipulate the system's dynamics via diffusion and decoherence. Incoherent control in this approach can be supplied by standard coherent control. In~\cite{PechenPRA2006}, two general classes of master equations were studied for quantum control, which correspond to weak coupling limit (WCL) and collisional based low density limit (LDL)~\cite{DumckeCMP1985, AccardiBook2002, AccardiJPA2002, PechenJMP2004, PechenJMP2006} in the theory of open quantum systems. Higher order corrections to the WCL~\cite{PechenIDAQP2002} can be included when necessary. Diffusion in collision-free models of quantum particles is also studied, e.g. in a quantum Poincar\'e model realizing behavior of ideal gas of noninteracting quantum Bolztman particles~\cite{KozlovSmolyanov2007}. The articles~\cite{PechenPRA2006, PechenPRA2011} contain 
the general theory for $N$-level quantum systems
and examples for the cases with $N=2, 4$. The work~\cite{PechenPRA2011} contains fundamental results about approximate
controllability of $N$-level quantum systems, whose dynamics is determined by coherent and incoherent controls. In~\cite{LokutsievskiyJPA2021}, an analytical description of reachable sets for two-level systems driven by coherent and incoherent controls was obtained using geometric control theory and surprisingly, unreachable states in the Bloch ball were discovered.
For the case of coherent and incoherent control of a two-level open quantum system, various quantum control problems were studied (generation of a target density matrix, including in minimal time,
maximizing Hilbert--Schmidt scalar product between final and target
density matrices, maximizing the Uhlmann--Jozsa fidelity,
estimating reachable and controllability sets, etc.) for various classes of controls and using various optimality conditions and optimization methods~\cite{MorzhinIJTP2019-2021, MorzhinLJM2019, MorzhinPhysPartNucl2020,MorzhinProcSteklov2021, LokutsievskiyJPA2021}. Quantum coherent and incoherent control of other systems has been studied, such as two-qubit systems (e.g.~\cite{PetruhanovIJMP2022}) and quantum harmonic oscillator~\cite{PechenChaos2022}. A model of incoherent control for the optical signals using the physics of a quantum heat engine was proposed~\cite{Qutubuddin2021}.

A particular result of~\cite{PechenPRA2011} is a two-stage method for approximate generation of a given target density matrix which was proposed and developed for generic $N$-level quantum systems and has analytical form of incoherent control on the first stage. At the first stage, coherent control is zero and constant in time incoherent control
is applied which is analytically computed  using eigenvalues of the target density matrix; this incoherent control drives the system state to a density matrix which approximately coincides 
with the intermediate target density matrix~$\widetilde{\rho}_{\rm target}$ whose eigenvalues are equal to eigenvalues of the target density 
matrix~$\rho_{\rm target}$. At the second stage, incoherent control is zero and optimized 
coherent control is applied to obtain final density matrix~$\rho(T)$
that with some sufficient accuracy coincides with~$\rho_{\rm target}$. This method allows to approximately create arbitrary density matrices of generic $N$-level quantum systems, but it does not consider time optimality, that motivates the importance of the problem of minimizing time necessary for the first stage which is typically most time consuming. In this regard, in Sec.~\ref{section4} of this article, we propose and compare with the original case the following modification of the first stage of the two-stage method for two-level systems:
instead of using constant incoherent control, optimization in the class of piecewise constant incoherent controls is performed. For this modification, we give exact analytical formulas for the quantum system's state 
at the end of the first stage, the objective functions, and their gradients depending on the parameters defining piecewise constant controls. Then we use these formulas for implementation of the two-step gradient projection method (GPM-2) --- projection version~\cite{AntipinDiffEq1994}
of the heavy ball method~\cite{PolyakZVMMF1964, PolyakBook1987}. 

\section{Formulation of the optimal control problems}
\label{section2}
 
The article \cite{VolkovJPA2021} considers the problem
of generating the phase shift quantum gate for a~two-level quantum system whose dynamics is determined by the Schr\"{o}dinger equation for evolution operator $U(t)$ and with objective functional $J_W$,
\begin{eqnarray}
	\label{Shrodinger_eq}
	\dfrac{dU(t)}{dt} &=& -i \left(H_0 + v(t) V \right) U(t), \qquad U(0) = \mathbb{I}_2, \qquad t \in [0, T], \\
	\label{obj_criterion_W}
	J_W(v) &=& \frac{1}{4} \left| {\rm Tr}(W^{\dagger} U(T)) \right|^2 \to\max, \qquad W = e^{i \varphi_W \sigma_z},
\end{eqnarray} 
where $U(t)$ and $W$ are $2 \times 2$ unitary matrices; $H_0$, $V$ are $2 \times 2$ Hermitian matrices; $v$ is coherent control being, in general, a measurable function;
$i$ is the imaginary unit. We consider the system of units where Planck's constant is $\hbar=1$. 
The goal is to find a unitary operator (matrix)~$U(T)$ for some~$T$ such that 
$U(T)$ coincides with $W$ or as close as possible to~$W$. The parameter~$\varphi_W$, which enters
in~$W$, and final time~$T$ allow to parameterize the optimal control problem.
The work~\cite{VolkovJPA2021} considers various domains 
on the coordinate plane of  $\varphi_W$ and~$T$. In this article, we consider the set 
$D := \left\{ (\varphi_W, T)~:~ \varphi_W \in \left(0, \frac{\pi}{2}\right), \quad
T \in \left(0, \frac{\pi}{2} \right] \right\}$.

The articles~\cite{PechenPRA2006, PechenPRA2011} contain the general theory
of controlling open $N$-level quantum systems driven by coherent and incoherent controls, 
and examples for $N=2, 4$. In~\cite{MorzhinIJTP2019-2021}, 
the case~$N=2$ is considered with piecewise continuous coherent~$v$
and incoherent~$n$ controls. The master equation for the density matrix $\rho(t)$, which is $2 \times 2$ Hermitian positive semidefinite matrix with unit trace ($\rho(t) = \rho^{\dagger}(t) \geq 0$, ${\rm Tr}\rho(t) = 1$), has the form:
\begin{eqnarray} 
	\frac{d \rho(t)}{dt} =
	-i[ H_0 + v(t) V, \rho(t)] + 
	\mathcal{L}_{n(t)}(\rho(t)), 
	\qquad \rho(t) \in \mathbb{C}^{2 \times 2}, \qquad \rho(0) = \rho_0.
	\label{quantum_system_rho}  
\end{eqnarray} 
The initial
density matrix~$\rho_0$ is fixed (in contrast, when studying controllability sets~\cite{MorzhinProcSteklov2021}, $\rho_0$ is not fixed). The dissipation superoperator $\mathcal{L}_{n(t)}$ acts on the density matrix as 
\begin{eqnarray} 
	\mathcal{L}_{n(t)} (\rho(t)) &=&
	\gamma \left( n(t) + 1 \right) \left( \sigma^- \rho(t) \sigma^+ - \dfrac{1}{2} \left\{ \sigma^+ \sigma^-, \rho(t) \right\} \right) \nonumber \\
	&& + \gamma n(t) \left( \sigma^+ \rho(t) \sigma^- - \dfrac{1}{2} \left\{ \sigma^- \sigma^+, \rho(t) \right\} \right), \qquad \gamma > 0 
\end{eqnarray}
and describes the controlled interactions between the system and its environment
(reservoir), where  control~$n$ is scalar. The notations $[A,B]= AB - BA$ and $\{A,B\}=AB + BA$ mean commutator and  anticommutator of operators $A,B$; matrices $\sigma^+ = \begin{pmatrix}
	0 & 0 \\ 1 & 0
\end{pmatrix}$, $\sigma^- = \begin{pmatrix}
	0 & 1 \\ 0 & 0
\end{pmatrix}$ determine the transitions between two energy levels,
control~$n$ is used to regulate this interaction. The following constraint is obligatory by the physical meaning of incoherent control:
\begin{eqnarray}
	n(t) \geq 0 \qquad \forall t \geq 0.
	\label{lower_bound_for_n}
\end{eqnarray}

In the closed (\ref{Shrodinger_eq}) and open (\ref{quantum_system_rho}) systems, 
the following forms of free Hamiltonian~$H_0$ and interaction Hamiltonian~$V$ are considered:
\begin{itemize}
	\item in the system (\ref{Shrodinger_eq}): $H_0 = \sigma_z$ and $V = \alpha_x \sigma_x + \alpha_y \sigma_y$; 
	\item in the system (\ref{quantum_system_rho}): $H_0 = \omega \begin{pmatrix}
		0 & 0 \\
		0 & 1
	\end{pmatrix} = \frac{\omega}{2}\left(\mathbb{I}_2 - \sigma_z \right)$ and
	$V = \mu \sigma_x$, where $\omega > 0$, $\mu \in \mathbb{R}$, $\mu \neq 0$. 
\end{itemize}
Here the Pauli matrices 
$\sigma_x = \begin{pmatrix}
	0 & 1 \\
	1 & 0
\end{pmatrix}$, $\sigma_y = \begin{pmatrix}
	0 & -i \\
	i & 0
\end{pmatrix}$, $\sigma_z = \begin{pmatrix}
	1 & 0 \\
	0 & -1
\end{pmatrix}$. We set $\alpha_x = \cos\theta$ and $\alpha_y = \sin\theta$.  
As noted in~\cite[Lemma 5]{PechenIzvRANSerMatem2016}, one can take $\theta = 0$ in $V$, 
taking into account the invariance of~$J_W$ with respect to $\theta$.  
Then $\alpha_x = 1$, $\alpha_y = 0$, and $V = \sigma_x$.
Thus, in the closed and open systems under consideration, 
the free Hamiltonians differ from each other additively by physically unrelevant $\mathbb{I}_2$ 
and by the scale factor (which is equivalent to time rescaling), while the interaction Hamiltonians, taking into account $\theta = 0$, 
are almost the same, they differ by the factor~$\mu$ that is equivalent to rescaling the amplitude of coherent control. Hence, up to rescaling time and amplitude of coherent control, they are the basically same. If we set $\gamma= 0$, then the system (\ref{quantum_system_rho}) becomes a closed system whose dynamics is described by the von Neumann equation (as in~\cite[Ch. IV]{ButkovskiyBook1990}).
Moreover, taking into account the relation $\rho(t) = U(t) \rho_0 U^{\dagger}(t)$, the system (\ref{quantum_system_rho}) for $\gamma=0$ can be described by (\ref{Shrodinger_eq}) with the corresponding Hamiltonian. 
This article considers the following classes of controls. In~Sec.~\ref{section3}, for the closed system~(\ref{Shrodinger_eq}), we consider the class of piecewise continuous controls when discussing the stationary point of the objective functional~$J_W$ and the PMP, and then we consider the following class of piecewise constant controls for reduction to a finite-dimensional optimization problem: 
\begin{eqnarray} 
	v(t) = \sum\limits_{k=1}^N a_k \chi_{[t_k, t_{k+1})}(t), \quad v(T) = v(T-), \quad \Delta t = T/N,
	\label{piecewise_constant_coherent_control_close_system}
\end{eqnarray} 
where $a_k \in \mathbb{R}$ is parameter representing control~$v$ during a partial interval $[t_k, t_{k+1})$ (``shelf''); $\chi_{[t_k, t_{k+1})}$ is the 
characteristic function of this interval; moreover, as an option, the constraint 
$|a_k| \leq \nu$ with some $\nu > 0$ can be used. 
In~Sec.~\ref{section4} for the open system~(\ref{quantum_system_rho}), when we formulate the modification of the first stage of the two-stage method the coherent control is~$v=0$ and incoherent control is  piecewise constant
\begin{equation}
	\label{piecewise_constant_incoherent_control_1st_stage_open_system}
	n(t) = \sum\limits_{k=1}^N a_k \chi_{[t_k,t_{k+1})}(t), \qquad a_k \geq 0, \qquad \Delta t = \widehat{t}/N,
\end{equation}
where $\widehat{t} > 0$ is the end time of the first stage. Here the constraint~(\ref{lower_bound_for_n}) is taken into account, and there may be the additional constraint $a_k \leq n_{\max}$, where $n_{\max} > 0$ is some given
value. In~Sec.~\ref{section4}, to implement the second stage with zero incoherent control, we consider (by analogy with the article~\cite{PechenPRA2011}) coherent control of the form 
\begin{eqnarray}
	\label{cos_coherent_control_2nd_stage_open_system}
	v(t) = A \cos(\Omega t), \quad t \in [\widehat{t}, T],
\end{eqnarray}
where $A$ is amplitude and~$\Omega$ is frequency. Moreover, in addition to the steering problem, we consider the condition $v(\widehat{t}) = v(T) = 0$ for coherent control~$v$, i.e. smooth switching on and off, correspondingly, at the moments $\widehat{t}$ and~$T$. In this regard, we consider the following class of controls:
\begin{eqnarray}
\label{zero_ends_sin_coherent_control_2nd_stage_open_system}
v(t) = A \sin\frac{\pi d (t - \widehat{t})}{T - \widehat{t}}, \quad 
d \in \{ 1, 2, \dots, \overline{d}\}, \quad 
T > \widehat{t}, \quad t \in [\widehat{t}, T], \quad \overline{d} \in \mathbb{N}.
\end{eqnarray}

In Sec.~\ref{section4}, we present a version of the two-stage method. At the first stage, we consider two variants of the problem of minimizing the squared Hilbert--Schmidt distance between~$\rho(\widehat{t})$ and
$\widetilde{\rho}_{\rm target}$: 
\begin{itemize}
	\item with a fixed moment $\widehat{t}$:
	\begin{eqnarray}
		J_1(n) = \| \rho(\widehat{t}) - \widetilde{\rho}_{\rm target} \|^2 \to \inf;
		\label{J_1_inf}
	\end{eqnarray}    
	\item with non-fixed moment $\widehat{t}$:
	\begin{eqnarray}
		\Phi(\widehat{t}, n; P) = \widehat{t} + P \| \rho(\widehat{t}) - \widetilde{\rho}_{\rm target} \|^2 \to \inf,  
		\label{time_minimal_control_problem_composite_Bloch_param}  
	\end{eqnarray}
	with some fixed penalty factor~$P>0$. 
\end{itemize}
At the second stage of the method, the system (\ref{quantum_system_rho}) is considered
with the initial condition $\rho(\widehat{t}) = \widehat{\rho}$, where $\widehat{\rho}$ is the density matrix obtained at the end of the first stage (it should be close enough to $\widetilde{\rho}_{\rm target}$).

\section{Nonoptimality of zero control satisfying the first order necessary optimality conditions for the quantum gate generation}
\label{section3}

\subsection{Realification of $U(t)$, satisfying the first order 
	necessary optimality conditions at the control $v = 0$} 

Various equivalent representations can be used here. Using the representation $U(t) = \begin{pmatrix} 
x_1(t) + i x_2(t) & x_3(t) + i x_4(t) \\
-x_3(t) + i x_4(t) & x_1(t) - i x_2(t)
\end{pmatrix}$, $x_j(t) \in \mathbb{R}$, $j = \overline{1,4}$ (by analogy with~\cite{GaronPRA2013}), the system~(\ref{Shrodinger_eq}) with $H_0 = \sigma_z$ and $V = \sigma_x$ is rewritten as the bilinear system
\begin{eqnarray}
	\label{Shrodinger_eq_in_terms_x}
	\dot x(t) = \left(A + B v(t)\right)x(t), \qquad x(0) = (1,0,0,0)^{\rm T}.
\end{eqnarray} 
Here $A = \begin{pmatrix}
A_2 & 0_2 \\
0_2 & A_2
\end{pmatrix}$, 
$B = \begin{pmatrix} 
0_2 & B_2 \\
-B_2 & 0_2
\end{pmatrix}$, where $A_2 = \begin{pmatrix}
0 & 1\\
-1 & 0
\end{pmatrix}$, 
$B_2 = \begin{pmatrix}
0 & 1 \\
1 & 0
\end{pmatrix}$; $0_2$ is $2 \times 2$ zero matrix; ``${\rm T}$'' 
means transpose. The objective functional~$J_W$~(\ref{obj_criterion_W}) to be maximized is rewritten as 
\begin{eqnarray}
	\label{obj_criterion_W_in_terms_x}
	J_W(v) = \left\langle x(T), L x(T) \right\rangle = \left( x_1(T) \cos\varphi_W + x_2(T) \sin\varphi_W \right)^2,
\end{eqnarray} 
where the final state $x(T)$ of the system~(\ref{Shrodinger_eq_in_terms_x}) 
is computed for a certain control~$v$; symmetric matrix $L = \begin{pmatrix}
L_2 & 0_2 \\
0_2 & 0_2
\end{pmatrix}$, where 
$L_2 = \begin{pmatrix}
\cos^2\varphi_W & \cos\varphi_W \sin\varphi_W \\
\cos\varphi_W \sin\varphi_W & \sin^2 \varphi_W
\end{pmatrix}$.

For this reformulation of the problem of maximizing the objective functional~$J_W$ in terms of
(\ref{obj_criterion_W_in_terms_x}) and (\ref{Shrodinger_eq_in_terms_x}), we consider the first-order  necessary optimality conditions known in the theory of optimal control~\cite{PontryaginBook1962, DemyanovBook1970, VasilievBook2011}. The Pontryagin function is $H(p,x,v) = \left\langle p, (A + Bv)x \right\rangle$,
where the variables $x,p \in \mathbb{R}^4$ and $v \in \mathbb{R}$, and the conjugate system 
\begin{eqnarray}
	\label{Schrodinger_Conjugate_system}
	\dot{p}(t) = -(A^{\rm T} + B^{\rm T} v(t)) p(t), \quad p(T) = 2Lx(T)
\end{eqnarray}
for some admissible process $(x, v)$. For any pair $(\varphi_W, T)$, the control $v = v^0 = 0$ satisfies the stationarity condition, which is a first-order necessary optimality condition
for piecewise continuous controls without constraints on their amplitudes: 
\begin{eqnarray}
	\label{Stationary_condition}
	\frac{\partial H(p,x,v)}{\partial v}\big|_{p=p^0(t), ~x=x^0(t)} = 
	\left\langle p^0(t), B x^0(t) \right\rangle \equiv 0, \quad t \in [0,T],
\end{eqnarray}
where $x^0$ is the solution of the system~(\ref{Shrodinger_eq_in_terms_x}) 
for $v=v^0$, and $p^0$ is the solution of the system~(\ref{Schrodinger_Conjugate_system}) 
for the process~$(x, v) = (x^0,v^0)$. If the pointwise constraint 
$v(t) \in Q := [-\nu, \nu]$ $\forall t \in [0, T]$ with some arbitrarily given $\nu>0$ is imposed, then we note that the pointwise maximum condition for the Pontryagin function in the PMP holds for any $\nu>0$ for the control $v^0 = 0$ in the class of piecewise continuous coherent controls (to the left of the ``$\equiv$'' sign, the letter ``$v$'' means variable $v \in \mathbb{R}$): 
\begin{eqnarray}
	\label{PMP}
	\max\limits_{v \in [-\nu, \nu]} H(p^0(t), x^0(t), v) \equiv H(p^0(t), x^0(t), v^0(t)), \quad t \in [0,T].
\end{eqnarray}

\subsection{Results of the finite-dimensional optimization} 

\begin{figure}[!ht]
	\centering
	\includegraphics[width=1\linewidth]{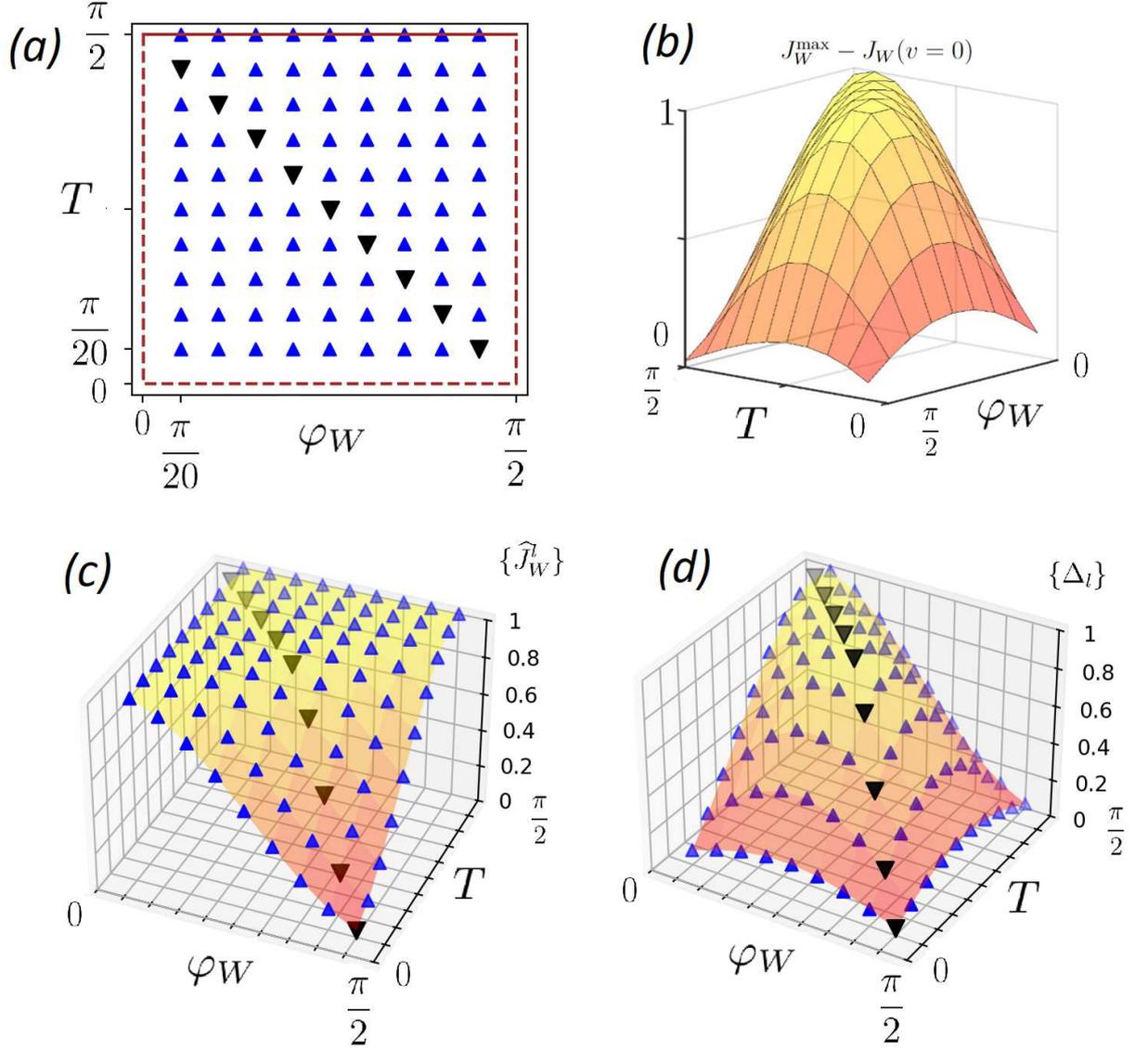} 
	\caption{Visualization of the numerical optimization results obtained for 	$(\varphi_W, T)\in\mathcal{D}$: 
		(a)~grid consisting of 90 nodes;
		(b)~(the GRAPE-type method's results) interpolation surface obtained based on the data from Table~1 showing the differences between the numerically maximized $J_W$ and $J_W(v=0)$ for each node of the grid~$\widehat{D}$; 
		(c)~(the stochastic methods' results) the table-defined function $\{\widehat{J}_W^l\}_{l=1}^{90}$ shown by markers, representing the computed maximal values of the objective functional~$J_W$, for each $l$-th node of the grid on $\mathcal{D}$, and the interpolation surface corresponding to this table-defined function;  
		(d)~(the stochastic methods' results) table-defined function $\{\Delta_l\}_{l=1}^{90} = \{\widehat{J}_W^l - J^l_W(v^0) \}_{l=1}^ {90}$, where $J^l_W(v^0)$ is the value of the objective functional for the control $v^0 = 0$, and the interpolation surface corresponding to this table-defined function.}
	\label{Figure1}
\end{figure}

In connection to the conditions (\ref{Stationary_condition}) and (\ref{PMP}), we numerically study whether the control $v^0 = 0$ is a global solution to the problem of maximizing the objective functional~$J_W$ for a given~$T$ in the class of piecewise constant controls of the form~(\ref{piecewise_constant_coherent_control_close_system}) for various pairs $\left(\varphi_W^j, T^i \right) \in D$, where the indices $j,i$ specify points in the grid $\widehat{D} := \left\{ \varphi_W^j = \frac{\pi}{20} j, \quad j = \overline{1,9}, \quad T_i = \frac{\pi }{20} i, \quad i = \overline{1,10} \right\} \subset D$ consisting of 90~nodes (see Figure~\ref{Figure1}(a)); all the nodes on the line $T = \frac{\pi}{2} - \varphi_W$ are marked with another marker. For piecewise constant control of the form~(\ref{piecewise_constant_coherent_control_close_system}), we denote ${\bf a}=(a_1,\dots,a_N)\in\mathbb R^N$.

For each node $(\varphi_W^j,~T^i) \in \widehat{D}$, the number $N$ of uniform partitions in $t$ is determined when defining piecewise constant controls of the form~(\ref{piecewise_constant_coherent_control_close_system}) so that 
$[0, T^i]$ is divided into $N$ equal parts. As in~\cite{VolkovJPA2021}, we use the following two approaches: 1)~GRAPE-type gradient search algorithm and exploiting exact analytically computed expression for gradient of the objective for piecewise constant controls which does not need solving differential evolution equation; 2)~dual annealing and differential evolution methods together.

For implementing GRAPE, we take $N=4+i$ and apply built in MATLAB function fminunc for unconstrainded optimization via quasi-Newton method using exact formula for gradient of the objective ($-J_W[{\bf a}]$). In numerical simulations, for each node we run algorithm starting with $10$ random initial controls where every control is produced by choosing  $a_k$ randomly and uniformly distributed in the interval $[-A;A]$ ($A=1$ is chosen), and then select the best result among 10.  The explicit expression for gradient of the objective used for optimization is~\cite{VolkovJPA2021}
\[
\nabla J_W[{\bf a}]=\frac{1}{2}\Im \left[{\rm Tr} (Y^\dagger){\rm Tr}(YV_{k})\right].
\]
Here
\begin{eqnarray*}
Y&=& W^\dagger U_T^{\bf a},\\
U^{\bf a}_T&=&U_N\dots U_k \dots U_1,\\
V_{k}&=&U^\dagger_k\dots U^\dagger_1 V U_1\dots U_k,\\
U_k&=&\cos\alpha_k-i\delta t(\sigma_z+a_k\sigma_x)\frac{\sin\alpha_k}{\alpha_k},\quad \alpha_k=\delta t\sqrt{1+a_k^2},\quad \delta t=\frac{T}{N}\,.
\end{eqnarray*}
For each run, maximum number of iterations allowed is $10^{6}$, maximum number of function evaluations allowed is $10^{6}$, and algorithm stops if $\|\nabla J_W\|_\infty\equiv \max\limits_i|(\nabla J_W)_i|< 10^{-8}$.  The results of optimization with the same stopping criteria are shown in Table~1 and Figure~\ref{Figure1}(b). Gradient based search algorithms are among the most suitable methods for the analysis of quantum control landscape, especially in the case they experience some difficulties that may indicate some fine details of the underlying control landscape. 

\begin{table}[h!] 
	\caption{For the quantum gate generation problem: computed using analytical expression for gradient differences between numerically maximized 
	$J_W$ and $J_W(v=0)$, i.e. $J_W^{\max} - J_W(v=0)$, for the grid $\widehat{D}$.}  
	\footnotesize
	\begin{tabular}{|l|l|l|l|l|l|l|l|l|l|}
		\hline
		\diagbox[width=5em]{$T_i$}{$\varphi_W^j$} & $\varphi_W^1$ & $\varphi_W^2$ & $\varphi_W^3$ & $\varphi_W^4$ & $\varphi_W^5$ & $\varphi_W^6$ & $\varphi_W^7$ & $\varphi_W^8$ & $\varphi_W^9$ \\ 
		\hline
		$T_1 = \pi/20$ &  0.092 &  0.161  & 0.214  & 0.246  &  0.254 & 0.237 & 0.197 & 0.138 & 0.066 \\ 
		\hline
		$T_2$ &  0.206 &  0.336 & 0.438  & 0.496 &  0.506 &  0.467  & 0.382  & 0.261 & 0.114 \\  
			\hline
		$T_3$ &  0.345 &  0.500 & 0.640 & 0.718 & 0.726 & 0.663 & 0.538 & 0.358 & 0.146 \\   
			\hline
		$T_4$ & 0.500 & 0.655 & 0.794 & 0.884  &  0.888  & 0.808  & 0.647 & 0.424 & 0.163 \\  
			\hline
		$T_5$ & 0.655 & 0.794 & 0.905 &  0.976 &  0.976 &  0.881 & 0.701 &  0.453 & 0.165 \\  
			\hline
		$T_6$ & 0.794 & 0.905 & 0.976 & \textbf{1.000} & 0.975 & 0.877 & 0.693 & 0.443 & 0.155 \\   
			\hline
		$T_7$ & 0.905 &  0.976 & \textbf{1.000} & 0.976 & 0.905 & 0.793 & 0.624 & 0.395 & 0.133 \\  
			\hline
		$T_8$ & 0.976 & \textbf{1.000} & 0.976 & 0.905 & 0.794 & 0.655 & 0.499 & 0.313 & 0.102 \\  
			\hline
		$T_9$ & \textbf{1.000} & 0.976 & 0.905 & 0.794 & 0.655 & 0.500 & 0.345 & 0.205 & 0.064 \\  
			\hline
		$T_{10} = \pi/2$ & 0.976 & 0.905 & 0.794 & 0.655 & 0.500 & 0.345 & 0.206 & 0.095 & 0.024 \\  
			\hline
\end{tabular} 
\normalsize
\end{table} 

Second approach uses two runs of the differential evolution method and two runs of the dual annealing method to minimize the objective function $f({\bf a}) := - J_W(v)$, where ${\bf a} = (a_k)_{k=1}^N$, $|a_k| \leq 50$,  with comparison of their results; the system (\ref{Shrodinger_eq_in_terms_x}) is solved numerically here. The minimization problem is considered here instead of the maximization problem, because the SciPy implementations of these methods are designed for minimization.

In Figure~\ref{Figure1}(c), via the markers, the table-defined
function $\{\widehat{J}_W^l\}_{l=1}^{90}$, which represents the computed optimal values of the objective functional $J_W(v)$ to be maximized for each $l$-th node grid, and also the interpolation surface corresponding to this table-defined function are shown. Figure~\ref{Figure1}(d) shows the table-defined function $\{\Delta_l\}_{l=1}^{90} = \{\widehat{J}_W^l - J ^l_W(v^0) \}_{l=1}^{90}$, where $J^l_W(v^0)$ is the value of the objective functional on the control $v^0 = 0$, and also the interpolation surface corresponding to this table-defined function. As is known (formula~(6) in~\cite{VolkovJPA2021}), $J_W(v^0) = \cos^2(\varphi_W + T)$. According to the computations, we have $\min\limits_{1 \leq l \leq 90} \{ \widehat{J}_W^l \}_{l=1}^{90} \approx 0.036$, $\max\limits_{1 \leq l \leq 90} \{ \widehat{J}_W^l \}_{l=1}^{90} \approx 1$, 
$\frac{1}{90} \sum\limits_{l=1}^{90} \widehat{J}_W^l \approx 0.843$ (see  Figure~\ref{Figure1}(c)); we also have $\min\limits_{1 \leq l \leq 90} \{ \Delta_l \}_{l=1}^{90} \approx 0.024$, $\max\limits_{1 \leq l \leq 90} \{ \Delta_l \}_{l=1}^{90} \approx 1$, 
$\frac{1}{90} \sum\limits_{l=1}^{90} \Delta_l \approx 0.565$ (see Figure~\ref{Figure1}(d)). 

Thus, it is shown that after reduction to the class of piecewise constant controls~(\ref{piecewise_constant_coherent_control_close_system}), 
the finite-dimensional optimization (GRAPE and stochastic methods) 
gives such piecewise constant controls 
for different pairs $\left(\varphi_W^j, T^i \right) \in D$ that the values 
of the objective functional $J_W(v)$ are substantially larger than for 
the control $v = v^0 = 0$. Since the class of piecewise constant controls
of the form~(\ref{piecewise_constant_coherent_control_close_system}) 
is nested in the class of piecewise continuous controls for the same $T$, 
then we have the situation, when such controls are found in the class 
of piecewise continuous controls that are significantly better in terms 
of the values of $J_W(v)$ than for the control $v = v^0 = 0$ satisfying the 
stationarity condition. In other words, the control $v = v^0 = 0$ is not optimal here. 
In addition, numerical results show that, for some part of the nodes
of the grid introduced above, we have $J_W$ equal to~1, i.e. there is full generation of the quantum
gate here. Note that in~\cite{PechenIzvRANSerMatem2016} and \cite[Theorem 3]{VolkovJPA2021} for a general class of measurable controls, it was shown that the control $v = v^0 = 0$ is a saddle point of the objective functional $J_W(v)$ for $(\varphi_W, T) \in D \setminus \left\{ (\varphi_W, T)~:~ \varphi_W + T \neq \pi/2 \right\}$ (the Hessian of the functional $J_W$ has negative and positive eigenvalues).

For comparison, note the article~\cite{ZobovJETP2014}, where, via an example of direct and inverse quantum Fourier transforms gates for qutrit with unitary dynamics (i.e. for a 3-level closed quantum system), a relationship between the phase factor of a quantum gate and the arrangement of energy levels of the effective Hamiltonian was shown.

\section{Modification of the first stage of the two-stage method for approximate steering}
\label{section4}

\subsection{Reduction to real states ($x(t)$), and objective functions in terms of intermediate final state $x(\widehat{t})$} 

Following the two-level system example given in the article~\cite{PechenPRA2011}, 
the article~\cite{MorzhinIJTP2019-2021} uses the Bloch parameterization 
$\rho = \dfrac{1}{2} \left( \mathbb{I}_2 + \sum\limits_{j \in \{x, y, z\}} x_j \sigma_j \right) =
\dfrac{1}{2} \begin{pmatrix}
	1 + x_3 & x_1 - i x_2 \\ x_1 + i x_2 & 1 - x_3
\end{pmatrix}$. 
Here $x = (x_1, x_2, x_3) \in \mathcal{B} := \{ x \in \mathbb{R}^3 : \| x \| \leq 1 \}$ is a Bloch vector representing some point in the Bloch ball; $\sigma_x$, $\sigma_y$ and $\sigma_z$ are the Pauli matrices; 
$x_j = {\rm Tr}(\rho \sigma_j)$, $j \in \{x,y,z\}$. The initial point $x_0$ has the coordinates 
$x_{0,j} = {\rm Tr} \left( \rho_0 \sigma_j \right)$, $j \in \{x,y,z\}$. Due to the bijection between density matrices and Bloch vectors, the Bloch parametrization allows to carry out the study in terms of the equivalent dynamical control system whose state $x(t)$ at time $t$ is the Bloch vector. Various equivalent parametrizations of quantum systems can be used~\cite{KozlovSmolyanov2021}. The following system is equivalent to~(\ref{quantum_system_rho}):
\begin{eqnarray} 
	\frac{dx(t)}{dt} &=& \left(A + B^v v(t) + B^n n(t) \right)x(t) + d, 
	\quad x(0) = x_0 \in \mathcal{B}, 
	\label{QS_Bloch_parametrization}
\end{eqnarray}
where $A = \begin{pmatrix}
	-\frac{\gamma}{2} & \omega & 0 \\
	-\omega & -\frac{\gamma}{2} & 0 \\
	0 & 0 & -\gamma
\end{pmatrix}$, 
$B^v = \begin{pmatrix}
	0 & 0 & 0 \\
	0 & 0 & -2 \mu \\
	0 & 2\mu & 0
\end{pmatrix}$, 
$B^n = \begin{pmatrix}
	-\gamma & 0 & 0 \\
	0 & -\gamma & 0 \\
	0 & 0 & -2\gamma
\end{pmatrix}$, 
$d = \begin{pmatrix}
	0 \\
	0 \\
	\gamma
\end{pmatrix}$. 

In terms of the Bloch parametrization and taking into account the parameterization (\ref{piecewise_constant_incoherent_control_1st_stage_open_system}) of coherent control, the following problems of finite-dimensional constrained optimization are considered in accordance with~(\ref{J_1_inf}) and~(\ref{time_minimal_control_problem_composite_Bloch_param}):  
\begin{itemize}
	\item with fixed moment $\widehat{t}$:
	\begin{eqnarray}
		g_1({\bf a}) := 2 J_1(n) = 
		\left\| x(\widehat{t}) - \widetilde{x}_{\rm target} \right\|^2 = 
		\sum\limits_{j=1}^3 \left( x_j(\widehat{t}) - \widetilde{x}_{{\rm target},j} \right)^2 \to \inf;
		\label{g1_inf}
	\end{eqnarray}     
	\item with non-fixed moment $\widehat{t}$:
	\begin{eqnarray}
		g_{\Phi}(\widehat{t}, {\bf a}; P') := 
		\Phi(\widehat{t}, n; P') = \widehat{t} + P' \| x(\widehat{t}) - \widetilde{x}_{\rm target} \|^2 \to \inf
		\label{gPhi_time_minimal_control_problem_composite_Bloch_param}  
	\end{eqnarray}
	with some penalty coefficient~$P' = P/2$. 
\end{itemize}
Here $\widetilde{\rho}_{{\rm target},j} = {\rm Tr}(\widetilde{\rho}_{\rm target} \sigma_j)$, $j \in \{x, y, z\}$. 

\subsection{Exact analytical formulas for the system's intermediate final state, objective functions and their gradients}

Below we show that the evolution equation (\ref{QS_Bloch_parametrization}) can be solved exactly for any initial state from the Bloch ball, any piecewise constant control~$n$ under zero control~$v$. This exact solution is used to represent the state $x(\widehat{t})$ as some function of the parameters $\{a_k\}$, which determine the control~$n$, and the moment $\widehat{t}$, if it is not fixed. In addition, the exact analytical formulas are obtained that explicitly show how the objective functions considered in the problems (\ref{g1_inf}) and (\ref{gPhi_time_minimal_control_problem_composite_Bloch_param}) depend on their arguments.

\begin{theorem} 
	\label{theorem_1}
	For a non-fixed or fixed final time $\widehat{t}$ and a vector ${\bf a}$ that define an incoherent control of the form (\ref{piecewise_constant_incoherent_control_1st_stage_open_system}) at the first stage of the method, the corresponding final state of the system (\ref{QS_Bloch_parametrization}) is $x(\widehat{t}, {\bf a}) := \left(x_1(\widehat{t}, {\bf a}),  x_2(\widehat{t}, {\bf a}), x_3(\widehat{t},  {\bf a}) \right)$, where 
	\begin{eqnarray}
		x_1(\widehat{t},{\bf a}) &:=&  Z(\widehat{t}, {\bf a}) \left(x_1(0) \cos(\omega \widehat{t}) + x_2(0) \sin(\omega \widehat{t}) \right), 
		\label{G_1_function}\\
		x_2(\widehat{t}, {\bf a}) &:=&  Z(\widehat{t}, {\bf a}) \left(x_2(0) \cos(\omega \widehat{t}) - x_1(0) \sin(\omega \widehat{t}) \right), 
		\label{G_2_function}\\
		&& Z(\widehat{t}, {\bf a}) = \exp\left( -\gamma \widehat{t} \left( \frac{1}{2} + \frac{1}{N} \sum\limits_{s=1}^N a_s \right) \right),
		\label{Z_function} \\
		x_3(\widehat{t}, {\bf a}) &:=&  x_3(0) \exp\left(-\gamma \widehat{t} \left( 1 + \frac{2}{N} \sum\limits_{s=1}^N a_s \right) \right) \nonumber \\
		&& + \sum\limits_{s=1}^{N-1} \frac{\left(1 - \exp\left(-\frac{\gamma \widehat{t}}{N} (1 + 2 a_s) \right) \right) } {1 + 2 a_s} \exp\left(-\frac{\gamma \widehat{t}}{N} \sum\limits_{m = s + 1}^N (1 + 2 a_m) \right) \nonumber \\
		&& + 
		\frac{1 - \exp\left(-\frac{\gamma \widehat{t}}{N} (1 + 2 a_N) \right)} {1 + 2 a_N}.
		\label{G_3_function}
	\end{eqnarray} 
\end{theorem}

The Theorem is proved in the next Subsection. Due to this Theorem we do not need to numerically solve the system (\ref{QS_Bloch_parametrization}).

\begin{corollary}
\label{corollary1}
If ${\bf a} = (p, \dots, p)$ with some $p \ge 0$ then $x_i$ in Theorem~\ref{theorem_1} are
\begin{eqnarray}
x_1(\widehat{t},{\bf a}) &=& Z(\widehat{t},{\bf a}) \left( x_1(0) \cos(\omega \widehat{t}) + x_2(0) \sin(\omega \widehat{t})\right), 
	\label{G1_when_control_n_is_p} \\
x_2(\widehat{t},{\bf a}) &=& Z(\widehat{t},{\bf a}) \left( x_2(0) \cos(\omega \widehat{t}) - x_1(0) \sin(\omega \widehat{t})\right),
	\label{G2_when_control_n_is_p} \\
x_3(\widehat{t},{\bf a}) &=& x_3(0) \exp\left(-\gamma \widehat{t} (1+2p) \right) + \frac{1 - \exp\left(-\gamma \widehat{t} (1+2p) \right)}{1 + 2p}.
	\label{G3_when_control_n_is_p}
\end{eqnarray} 
\end{corollary}

The formula (\ref{G3_when_control_n_is_p}) is obtained from 
(\ref{G_3_function}) using the factorization formula 
$1 - y^N = (1 - y) \sum\limits_{s=1}^N y^{N-s} = (1-y) \left(1 +  \sum\limits_{s=1}^{N-1} y^{N-s}\right)$. Moreover, in   Sec.~5 of the article~\cite{MorzhinProcSteklov2021}, such the case is considered that the control~$v=0$ and incoherent control~$n$ is constant over the entire time interval. Taking $t = \widehat{t}$ in the formula~(5.1) in~\cite{MorzhinProcSteklov2021}, we also get (\ref{G1_when_control_n_is_p})--(\ref{G3_when_control_n_is_p}).

\begin{theorem} \label{theorem_2}
	For the functions $x_j(\widehat{t}, {\bf a})$, $j=1,2,3$, which are defined in (\ref{G_1_function}),
	(\ref{G_2_function}) and (\ref{G_3_function}), their first order partial derivatives are as follows: 
	\begin{eqnarray}
		\frac{\partial x_1(\widehat{t}, {\bf a})}{\partial a_q} &=& -\frac{\gamma \widehat{t}}{N} Z(\widehat{t}, {\bf a}) 
		\left( x_1(0) \cos(\omega \widehat{t}) + x_2(0) \sin(\omega \widehat{t}) \right),  
		\label{G_1_funtion_derivative_a_q} \\
		\frac{\partial x_1(\widehat{t}, {\bf a})}{\partial \widehat{t}} &=& Z(\widehat{t}, {\bf a}) 
		\Big[ -\gamma \left(\frac{1}{2} + \frac{1}{N} \sum\limits_{s=1}^N a_s \right) \left( x_1(0) \cos(\omega \widehat{t}) + x_2(0) \sin(\omega \widehat{t}) \right) \nonumber \\
		&& -
		x_1(0) \omega \sin(\omega \widehat{t}) + x_2(0) \omega \cos(\omega \widehat{t}) \Big],
		\label{G_1_funtion_derivative_time} \\
		\frac{\partial x_2(\widehat{t}, {\bf a})}{\partial a_q} &=& -\frac{\gamma \widehat{t}}{N} Z(\widehat{t}, {\bf a}) 
		\left( x_2(0) \cos(\omega \widehat{t}) - x_1(0) \sin(\omega \widehat{t}) \right),  \\
		\frac{\partial x_2(\widehat{t}, {\bf a})}{\partial \widehat{t}} &=& Z(\widehat{t}, {\bf a}) 
		\Big[ -\gamma \left(\frac{1}{2} + \frac{1}{N} \sum\limits_{s=1}^N a_s \right) \left( x_2(0) \cos(\omega \widehat{t}) - x_1(0) \sin(\omega \widehat{t}) \right) \nonumber \\
		&& -
		x_2(0) \omega \sin(\omega \widehat{t}) - x_1(0) \omega \cos(\omega \widehat{t}) \Big],
		\label{G_2_funtion_derivative_time}
	\end{eqnarray}
	where $Z(\widehat{t}, {\bf a})$ is defined in (\ref{Z_function}), index $q = 1, 2, \dots, N$;
	\begin{eqnarray}
		\frac{\partial x_3(\widehat{t}, {\bf a})}{\partial a_q} &=&
		-\frac{2 \gamma \widehat{t}}{N} x_3(0)  \exp\left(-\gamma \widehat{t} \left(1 + \frac{2}{N} \sum\limits_{s=1}^N a_s \right) \right) + W(\widehat{t}, {\bf a}), 
		\label{G_3_derivative_a_q}
	\end{eqnarray}
	where
	\begin{eqnarray} 
		\label{W_function}
		W(\widehat{t}, {\bf a}) &:=& 
		\begin{cases}
			D(\widehat{t}, a_1) \exp\left(-\frac{\gamma \widehat{t}}{N} \sum\limits_{m=2}^N (1 + 2 a_m) \right), & \text{if} \quad q = 1, \\
			H(\widehat{t}, a, q) + D(\widehat{t}, a_q) \\ \times \exp\left(-\frac{\gamma \widehat{t}}{N} \sum\limits_{m=q+1}^N (1 + 2 a_m) \right), & \text{if} \quad 1 < q \leq N-1, \\
			H(\widehat{t}, a, N) + D(\widehat{t}, a_N), & \text{if} \quad q = N,  
		\end{cases} \\
		D(\widehat{t}, a_r) &:=& -\frac{2\left(1 - \exp\left( -\frac{\gamma \widehat{t}}{N} \left(1 + 2 a_r \right) \right) \right)}{(1 + 2 a_r)^2} + 
		\frac{2\gamma \widehat{t} \exp\left( - \frac{\gamma \widehat{t}}{N} (1 + 2 a_r) \right)}{(1 + 2 a_r)N}, \\
		H(\widehat{t}, a, k) &:=& -\frac{2 \gamma \widehat{t}}{N} 
		\sum\limits_{s=1}^{k-1} \frac{1-\exp\left(-\frac{\gamma \widehat{t}}{N}(1+ 2 a_s) \right)}{1 + 2 a_s}  \exp\left(-\frac{\gamma \widehat{t}}{N} \sum\limits_{m=s+1}^N (1 + 2 a_m) \right), \qquad 
		\label{H_function}
	\end{eqnarray}
	index $q = 1, 2, \dots, N$;
	\begin{eqnarray}
		\frac{\partial x_3(\widehat{t}, {\bf a})}{\partial \widehat{t}} &=&
		-\frac{2\gamma}{N} x_3(0) \exp\left(-\gamma \widehat{t} \left(1 + \frac{2}{N} \sum\limits_{s=1}^N a_s \right) \right) \sum\limits_{s=1}^N a_s \nonumber \\
		&& + \sum\limits_{s=1}^{N-1} \frac{\gamma}{(1 + 2a_s)N} 
		\exp\left(-\frac{\gamma \widehat{t}}{N} \sum\limits_{m=s+1}^N (1 + 2 a_m) \right)\nonumber \\
		&& \times\Big[ (1 + 2a_s) \exp\Big(-\frac{\gamma \widehat{t}}{N} (1 + 2a_s) \Big)  \nonumber \\
		&& + \Big(\exp\Big( -\frac{\gamma \widehat{t}}{N} (1 + 2 a_s) \Big) - 1 \Big) \sum\limits_{m=s+1}^N(1+ 2a_m) \Big]  \nonumber \\ 
		&& +\frac{\gamma}{N} \exp\Big(-\frac{\gamma \widehat{t}}{N} (1 + 2 a_N) \Big),
		\label{G_3_function_derivative_T}
	\end{eqnarray}
	where $q = 1, 2, \dots, N$. For the objective functions $g_1({\bf a})$ and $g_{\bf \Phi}(\widehat{t}, {\bf a}; P')$ considered in the problems (\ref{g1_inf}) and (\ref{gPhi_time_minimal_control_problem_composite_Bloch_param}), their 
	first order partial derivatives in terms  (\ref{G_1_funtion_derivative_a_q})---(\ref{G_3_function_derivative_T}) are: 
	\begin{eqnarray}  
		\frac{\partial g_1({\bf a})}{\partial a_q} &=& 
		2 \left\langle x(\widehat{t}, {\bf a}) - \widetilde{x}_{\rm target}, \frac{\partial x(\widehat{t}, {\bf a})}{\partial a_q} \right\rangle \nonumber \\
		& =& 2 \sum\limits_{j=1}^3 \left(x_j(\widehat{t}, {\bf a}) - \widetilde{x}_{{\rm target},j} \right) \frac{\partial x_j(\widehat{t}, {\bf a})}{\partial a_q}, \quad q = 1, 2, \dots, N; \label{g1_derivative} \\
		\frac{\partial g_{\Phi}(\widehat{t}, {\bf a}; P')}{\partial a_q} &=& 
		2 P' \left\langle 
		x(\widehat{t}, {\bf a}) - \widetilde{x}_{\rm target}, 
		\frac{\partial x(\widehat{t}, {\bf a})}{\partial a_q} \right\rangle, \quad q = 1, 2, \dots, N; \label{gPhi_derivative_a_q} \\ 
		\frac{\partial g_{\Phi}(\widehat{t}, {\bf a}; P')}{\partial \widehat{t}} &=& 
		1 + 2 P' \left\langle 
		x(\widehat{t}, {\bf a}) - \widetilde{x}_{\rm target}, 
		\frac{\partial x(\widehat{t}, {\bf a})}{\partial \widehat{t}} \right\rangle. 
		\label{gPhi_derivative_time}
	\end{eqnarray} 
\end{theorem} 

\subsubsection{Proofs of Theorems 1, 2}

{\bf Proof of Theorem \ref{theorem_1}.} 
Consider the equation (\ref{QS_Bloch_parametrization}) sequentially on partial intervals of length $\Delta t = \widehat{t}/N$ with initial conditions $x(t_1=0)=x_0$, $x(t_k)=x(t_k-0)$, $k=2, 3, \dots, N$. For each $k$th initial condition, the corresponding Cauchy problem for the system of ordinary differential equations has the exact solution 
\begin{eqnarray*}
	x_1(t) &=& \exp\left(-\frac{\gamma}{2}(1+2a_k)(t-t_k) \right)\left(x_1(t_k)\cos(\omega(t-t_k)) + x_2(t_k)\sin(\omega(t-t_k)) \right), \\
	x_2(t) &=& \exp\left(-\frac{\gamma}{2}(1+2a_k)(t-t_k) \right)\left(x_2(t_k)\cos(\omega(t-t_k)) - x_1(t_k)\sin(\omega(t-t_k)) \right), \\
	x_3(t) &=& \exp\left(-\gamma (1+2a_k)(t-t_k) \right) x_3(t_k) + 
	\frac{1-\exp(-\gamma(1+2a_k)(t-t_k))}{1+2a_k}. 
\end{eqnarray*}
Thus, for $k = 1, 2, \dots, N$, we have
\begin{eqnarray}
	x_1(t_{k+1}) &=& E^{\rm I}(\widehat{t}, a_k) \left(x_1(t_k)\cos(\omega \Delta t) + x_2(t_k)\sin(\omega \Delta t) \right), \label{x1tkplus1} \\
	x_2(t_{k+1}) &=& E^{\rm I}(\widehat{t}, a_k) \left(x_2(t_k)\cos(\omega \Delta t) - x_1(t_k)\sin(\omega \Delta t) \right), \label{x2tkplus1} \\ 
	x_3(t_{k+1}) &=& E^{\rm II}(\widehat{t}, a_k) x_3(t_k) + \frac{1- E^{\rm II}(\widehat{t}, a_k)}{1+2a_k}, \label{x3tkplus1}  
\end{eqnarray}
where $E^{\rm I}(\widehat{t}, a_k) := \exp\left(- \frac{\gamma \widehat{t} (1+2a_k)}{2 N} \right)$ and
$E^{\rm II}(\widehat{t}, a_k) := \exp\left(-\frac{\gamma \widehat{t} (1+2a_k)}{N} \right)$. The equations  
(\ref{x1tkplus1})--(\ref{x3tkplus1}) are difference~\cite{RomankoBook2012}  and linear. With the initial conditions $x_j(t_1=0)=x_{x,j}$, $j=1,2,3$, consider the Cauchy problem for these difference equations. The equations (\ref{x1tkplus1}), (\ref{x2tkplus1}) do not depend  on (\ref{x3tkplus1}). If $k=1$, then (\ref{x1tkplus1}), (\ref{x2tkplus1}) give 
\begin{eqnarray*}
	x_1(t_2) &=& E^{\rm I}(\widehat{t}, a_1) \left(x_1(t_1=0)\cos(\omega \Delta t) + x_2(t_1=0)\sin(\omega \Delta t) \right), 
	\\
	x_2(t_2) &=& E^{\rm I}(\widehat{t}, a_1) \left(x_2(t_1=0)\cos(\omega \Delta t) - x_1(t_1=0)\sin(\omega \Delta t) \right). 
\end{eqnarray*}
Further, using (\ref{x1tkplus1}), (\ref{x2tkplus1}) sequentially for $k=2,3$  
and substituting the right-hand sides of the equations instead of
$x_j(t_k)$, $j=1,2$, we obtain the following non-recurrent 
(i.e. only with $x_j(0)$, $j=1,2$) formulas for $x_1(t_{k+1})$, $x_2(t_{k+1})$, $k = 2, 3$: 
\begin{eqnarray*}
	x_1(t_3) &=& E^{\rm I}(\widehat{t}, a_2) \left(x_1(t_2)\cos(\omega \Delta t) + x_2(t_2)\sin(\omega \Delta t) \right) 	
	\nonumber \\
	&=& E^{\rm I}(\widehat{t}, a_1) E^{\rm I}(\widehat{t}, a_2) \left(x_1(0)\cos(2\omega \Delta t) + x_2(0)\sin(2 \omega \Delta t) \right),   \\
	x_2(t_3) &=& E^{\rm I}(\widehat{t}, a_2) \left(x_2(t_2)\cos(\omega \Delta t) - x_1(t_2)\sin(\omega \Delta t) \right) \nonumber \\
	&=& E^{\rm I}(\widehat{t}, a_1) E^{\rm I}(\widehat{t}, a_2) \left(x_2(0)\cos(2\omega \Delta t) - x_1(0)\sin(2 \omega \Delta t) \right), \\
	x_1(t_4) &=& E^{\rm I}(\widehat{t}, a_3) \left(x_1(t_3)\cos(\omega \Delta t) + x_2(t_3)\sin(\omega \Delta t) \right) \nonumber \\
	&=& \prod\limits_{s=1}^3 E^{\rm I}(\widehat{t}, a_s) \left(x_1(0)\cos(3\omega \Delta t) + x_2(0)\sin(3 \omega \Delta t) \right),   \\
	x_2(t_4) &=& E^{\rm I}(\widehat{t}, a_3) \left(x_2(t_3)\cos(\omega \Delta t) - x_1(t_3)\sin(\omega \Delta t) \right) \nonumber \\ 
	&=& \prod\limits_{s=1}^3 E^{\rm I}(\widehat{t}, a_s) \left(x_2(0)\cos(3\omega \Delta t) - x_1(0)\sin(3 \omega \Delta t) \right).  
\end{eqnarray*}
Based on these results, we write
\begin{eqnarray}
	x_1(t_{k+1}) &=& \prod\limits_{s=1}^k E^{\rm I}(\widehat{t}, a_s) \left(x_1(0) \cos(k \omega \Delta t) + x_2(0) \sin(k \omega \Delta t) \right), 
	\label{x1tkplus1_nonrecurrent} \\
	x_2(t_{k+1}) &=& \prod\limits_{s=1}^k E^{\rm I}(\widehat{t}, a_s) \left(x_2(0) \cos(k \omega \Delta t) - x_1(0) \sin(k \omega \Delta t) \right), 
	\label{x2tkplus1_nonrecurrent}
\end{eqnarray}
where $k = 1, 2, \dots, N$. Using (\ref{x1tkplus1_nonrecurrent}), 
(\ref{x2tkplus1_nonrecurrent}) for $k=N$ with the condition
$N \omega \Delta t = \omega \widehat{t}$ and rewriting 
$\prod\limits_{s=1}^N E^{\rm I}(\widehat{t}, a_s) = \exp\left( -\gamma \widehat{t}\left(\frac{1}{2} + \frac{1}{N} \sum\limits_{s=1}^N a_s \right) \right)$, 
we obtain (\ref{G_1_function})--(\ref{Z_function}). 

The equation (\ref{x3tkplus1}) does not depend  on (\ref{x1tkplus1}), (\ref{x2tkplus1}). Using sequentially (\ref{x3tkplus1}) for $k = 1, 2, 3, 4$, we obtain
\begin{eqnarray*}
	x_3(t_2) &=& E^{\rm II}(\widehat{t}, a_1) x_3(t_1=0) + \frac{1 - E^{\rm II}(\widehat{t}, a_1)}{1 + 2a_1}, \\
	x_3(t_3) &=& E^{\rm II}(\widehat{t}, a_2) x_3(t_2) + \frac{1 - E^{\rm II}(\widehat{t}, a_2)}{1 + 2a_2} \nonumber \\
	&=& E^{\rm II}(\widehat{t}, a_1) E^{\rm II}(\widehat{t}, a_2) x_3(0) + \frac{E^{\rm II}(\widehat{t}, a_2)\left(1 - E^{\rm II}(\widehat{t}, a_1) \right)}{1 + 2a_1} +
	\frac{1 - E^{\rm II}(\widehat{t}, a_2)}{1 + 2a_2}, \\ 
	x_3(t_4) &=& E^{\rm II}(\widehat{t}, a_3) x_3(t_3) + \frac{1-E^{\rm II}(\widehat{t}, a_3)}{1+2a_3} \nonumber  \\
	&=& \prod\limits_{s=1}^3 E^{\rm II}(\widehat{t}, a_s) x_3(0) + \frac{E^{\rm II}(\widehat{t}, a_2) E^{\rm II}(\widehat{t}, a_3)\left( 1 - E^{\rm II}(\widehat{t}, a_1)\right)}{1 + 2a_1} \nonumber \\
	&&+ \frac{E^{\rm II}(\widehat{t}, a_3)\left( 1 - E^{\rm II}(\widehat{t}, a_2)\right)}{1 + 2a_2} + \frac{1 - E^{\rm II}(\widehat{t}, a_3)}{1 + 2a_3},\\ 
	x_3(t_5) &=&  E^{\rm II}(\widehat{t}, a_4) x_3(t_4) + \frac{1-E^{\rm II}(\widehat{t}, a_4)}{1 + 2a_4} \nonumber \\
	&=& \prod\limits_{s=1}^4 E^{\rm II}(\widehat{t}, a_s) x_3(0) + \frac{\prod\limits_{m=2}^4 E^{\rm II}(\widehat{t}, a_m)\left(1 - E^{\rm II}(\widehat{t}, a_1) \right)}{1 + 2a_1} \nonumber \\
	&& +\frac{\prod\limits_{m=3}^4 E^{\rm II}(\widehat{t}, a_m)\left(1 - E^{\rm II}(\widehat{t}, a_2) \right)}{1 + 2a_2} + 
	\frac{E^{\rm II}(\widehat{t}, a_4)\left(1 - E^{\rm II}(\widehat{t}, a_3) \right)}{1 + 2a_3} +  \frac{1 - E^{\rm II}(\widehat{t}, a_4)}{1 + 2a_4}.
\end{eqnarray*} 
Based on these relations, we write
\begin{eqnarray}
	x_3(t_{k+1}) &=& x_3(0) \prod\limits_{s=1}^k E^{\rm II}(\widehat{t}, a_s) \nonumber \\
	&& +\sum\limits_{s=1}^{k-1} \frac{\prod\limits_{m=s+1}^k E^{\rm II}(\widehat{t}, a_m)\left(1 - E^{\rm II}(\widehat{t}, a_s) \right)}{1 + 2 a_s}  + \frac{1 - E^{\rm II}(\widehat{t}, a_k)}{1 + 2 a_k}, 
	\label{x3tkplus1_nonrecurrent}
\end{eqnarray}
where $k = 1, 2, \dots, N$. Using (\ref{x3tkplus1_nonrecurrent}) for $k=N$ and 
transforming the products $\prod\limits_{s=1}^N E^{\rm II}(\widehat{t}, a_s)$ and 
$\prod\limits_{m=s+1}^N E^{\rm II}(\widehat{t}, a_m)$, we obtain (\ref{G_3_function}).

The proof is complete.

{\bf Proof of Theorem \ref{theorem_2}.} The derivatives (\ref{G_1_funtion_derivative_a_q})--(\ref{G_2_funtion_derivative_time}) and (\ref{G_3_function_derivative_T})--(\ref{gPhi_derivative_time})
are easily obtained by the rules of differentiation. To obtain the derivative (\ref{G_3_derivative_a_q})--(\ref{H_function}), we take into account the difference 
between the indices in the second term in (\ref{G_3_function}).
The proof is complete.

\subsubsection{Gradient projection method using the exact forms for the objective 	functions and their gradients} 

Consider such the set $Q$ that $Q = Q_{\infty} := [0, \infty)$ or 
$Q = Q_{n_{\max}} := [0, n_{\max}]$ with a given $n_{\max}>0$. 

In finite-dimensional unconstrained optimization, 
the heavy ball method~\cite{PolyakZVMMF1964, PolyakBook1987} is known. 
Also its projection version~\cite{AntipinDiffEq1994}, GPM-2, is known 
in finite-dimensional constrained optimization. 
GPM-2 (see the formula (3.3) in~\cite{AntipinDiffEq1994}) 
is adapted in the present work and has the following 
form as applied to the problem of minimizing the objective function 
$g_1({\bf a})$ taking into account~(\ref{g1_derivative}): 
\begin{eqnarray}
	\label{initial_vector_a}
	{\bf a}^{(0)} &\in& Q \quad \text{is a given admissible initial vector},\\
	\label{iteration_1_GPM}
	a_k^{(1)} &=& \overline{a}_k^{(0)}(\beta^{(0)}),
	\quad k =\overline{1,N}, \\
	\overline{a}_k^{(0)}(\beta) &:=&  
	{\rm Pr}_Q\left(a_k^{(0)} - 
	\beta \frac{\partial g_1({\bf a})}{\partial a_k}\Big|_{{\bf a} = {\bf a}^{(0)}} \right),\\
	a_k^{(m+1)} &=& \overline{a}_k^{(m)}(\beta^{(m)};\lambda), \quad k =\overline{1,N}, \quad m = \overline{1,M}, 
	\label{a_k_mplus1_computation} \\
	\overline{a}_k^{(m)}(\beta;\lambda) &:=& {\rm Pr}_Q \Big(a_k^{(m)} - 
	\beta  \frac{\partial g_1({\bf a})}{\partial a_k}\Big|_{{\bf a} = {\bf a}^{(m)}} + 
	\lambda \left(a_k^{(m)} - a_k^{(m-1)} \right) \Big),  \\
	{\rm Pr}_Q(z) &=& 
	\begin{cases}
		0, & \text{if} \quad z < 0, \\
		z, & \text{if} \quad Q = Q_{\infty}
		\quad \text{and} \quad z \in Q_{\infty},\\
		z, & \text{if} \quad Q = Q_{n_{\max}} \quad \text{and} \quad z \in Q_{\max},\\
		n_{\max}, & \text{if} \quad Q = Q_{\max} \quad \text{and} \quad z > n_{\max}, 
	\end{cases}
	\label{GPM_projection}
\end{eqnarray} 
where the parameters $\beta^{(m)}>0$  ($m=\overline{0,M}$) can, as a variant, be sought as fixed the same for all iterations and providing convergence of the method; also it is required to fix some appropriate value of the parameter  
$\lambda \in (0, 1)$  for the all iterations. As a condition for the end of computations,
consider the inequality $g_1({\bf a}) < \varepsilon$ with some sufficiently small $\varepsilon>0$ along with some ``ceiling'' for   numbers of such iterations. The first iteration uses the one-step GPM~\cite{PolyakBook1987}. 

By analogy with (\ref{initial_vector_a})--(\ref{GPM_projection}), 
we adapt GPM-2 for minimizing the objective function 
$g_{\Phi}(\widehat{t}, {\bf a}; P')$ taking into account~(\ref{gPhi_derivative_a_q}),~(\ref{gPhi_derivative_time}):
\begin{eqnarray*}
	\label{GPM_initial_pair_a_time}
	(\widehat{t}, {\bf a}^{(0)}) &\in& [\widehat{t}_{\min}, \widehat{t}_{\max}] \times Q   \quad \text{is a given admissible initial pair},\\
	a_k^{(1)} &=& \overline{a}_k^{(0)}(\beta^{(0)}), \quad k =\overline{1,N}, \\
	\overline{a}_k^{(1)}(\beta) &=& {\rm Pr}_Q\left(a_k^{(0)} - 
	\beta \frac{\partial g_{\Phi}(\widehat{t}, {\bf a}; P')}{\partial a_k}\Big|_{(\widehat{t}, {\bf a}) = (\widehat{t}^{(0)}, {\bf a}^{(0)})} \right), \\ 
	\widehat{t}^{(1)} &=& \overline{\widehat{t}}^{(0)}(\beta^{(0)}), 
	\label{tfinal_formula1} \\
	\label{tfinal_formula2}
	\overline{\widehat{t}}^{(0)}(\beta) &=&
	{\rm Pr}_{[\widehat{t}_{\min}, \widehat{t}_{\max}]}\left(\widehat{t}^{(0)} - 
	\beta \frac{\partial g_{\Phi}(\widehat{t}, {\bf a}; P')}{\partial \widehat{t}}\Big|_{(\widehat{t}, {\bf a}) = (\widehat{t}^{(0)}, {\bf a}^{(0)})} \right),\\
	a_k^{(m+1)} &=& \overline{a}_k^{(m)}(\beta^{(m)},\lambda), \quad k =\overline{1,N}, \quad m = \overline{1,M},\\
	\overline{a}_k^{(m)}(\beta,\lambda) &=& {\rm Pr}_Q \Big(a_k^{(m)} - 
	\beta \frac{\partial g_{\Phi}(\widehat{t}, {\bf a}; P')}{\partial a_k}\Big|_{(\widehat{t}, {\bf a}) = (\widehat{t}^{(0)}, {\bf a}^{(0)})} +   
	\lambda \left(a_k^{(m)} - a_k^{(m-1)} \right) \Big), \\
	\widehat{t}^{(m+1)} &=& \overline{\widehat{t}}^{(m)}(\beta^{(m)},\lambda), \quad m = \overline{1,M}, \\
	\overline{\widehat{t}}^{(m)}(\beta,\lambda) &=& 
	{\rm Pr}_{[\widehat{t}_{\min}, \widehat{t}_{\max}]} \Big(\widehat{t}^{(m)} - 
	\beta \frac{\partial g_{\Phi}(\widehat{t}, {\bf a}; P')}{\partial \widehat{t}}\Big|_{(\widehat{t}, {\bf a}) = (\widehat{t}^{(0)}, {\bf a}^{(0)})} + 
	\lambda \left(\widehat{t}^{(m)} - \widehat{t}^{(m-1)} \right) \Big), 
\end{eqnarray*}
where values $\widehat{t}_{\min}$ and $\widehat{t}_{\max}$ are fixed such that 
$0 < \widehat{t}_{\min} <  \widehat{t}_{\max}$; the parameters
$\beta^{(m)}>0$ ($m=\overline{0,M}$) should be adjusted.

Further, in addition, one can restrict variations of piecewise constant controls via the constraints $0 \leq (a_{k+1} - a_k)^2 \leq \delta_a$, $\delta_a > 0$, $k = 1, \dots, N-1$. To do this, consider an objective function that includes the function $g_1({\bf a})$ and the additive penalty function $R^{\alpha}({\bf a})$ by the method of external penalty functions~\cite[Ch. 9]{PolyakBook1987}: 
$g_1^{\alpha}({\bf a}) = g_1({\bf a}) + R^{\alpha}({\bf a}) \to \inf\limits_a$,
$R^{\alpha}({\bf a}) = \alpha \sum\limits_{k=1}^N \left(\max\{ (a_{k+1} - a_k)^2 - \delta_a, 0 \} \right)^2$, where the cutoff functions are squared for continuity of the partial derivatives, and the parameter $\alpha > 0$. The penalty function's gradient is formed from the partial derivatives $\frac{\partial R^{\alpha}({\bf a})}{\partial a_1} = 4 \max\{ (a_2 - a_1)^2 - \delta_a, 0 \}(a_1 - a_2)$, $\frac{\partial R^{\alpha}({\bf a})}{\partial a_N} = 4 \max\{ (a_N - a_{N-1})^2 - \delta_a, 0 \}(a_N - a_{N-1})$, $\frac{\partial R^{\alpha}({\bf a})}{\partial a_{k}} = 4 \max\{ (a_k - a_{k-1})^2 - \delta_a, 0 \}(a_k - a_{k-1}) + 4 \max\{ (a_{k+1} - a_k)^2 - \delta_a, 0 \}(a_k - a_{k+1})$, $1 < k < N$. 

For comparison, in the article \cite{MorzhinIJTP2019-2021} for the system (\ref{quantum_system_rho}), the time-minimal control problem was considered with the terminal constraint 
$\rho(T) = \rho_{\rm target}$. For this problem, the approach was considered with a series of optimal control problems without terminal
constraint, with minimizing the square of the Hilbert--Schmidt distance, i.e. $\|\rho(T_i) - \rho_{\rm target} \|^2 = {\rm Tr} \left( \rho(T_i) - \rho_{\rm target} \right)^2$, on descending sequences of fixed final times $\{T_i\}$ under piecewise continuous coherent and incoherent controls, which in the computer implementation of the algorithm were considered as piecewise linear functions. For each such auxiliary optimization problem, 
the two-parameter one-step GPM with the Fr\'{e}chet functional derivative 
was applied (for more general class of optimal control problems with free final state, 
the theory of optimal control gives more general formula for the functional 
derivative~\cite{DemyanovBook1970, VasilievBook2011}, which was specified
in~\cite{MorzhinIJTP2019-2021}).

\subsection{Comparing the results of the various first stages} 

The values $\omega = 1$,  $\gamma = 0.002$, $\mu = 0.01$, and $n_{\max} = 100$ are taken. 

{\bf Example 1}. Set the initial state $x_0 = (1, 0, 0)$, which corresponds to the initial density matrix 
$\rho_0 = \frac{1}{2} \begin{pmatrix}
	1 & 1\\
	1 & 1
\end{pmatrix}$, and the target state $x_{\rm target} = (0, 0, -\frac{1}{2})$, which
corresponds to the target density matrix
$\rho_{\rm target} = \begin{pmatrix}
	1/4 & 0\\
	0 & 3/4
\end{pmatrix}$. The eigenvalues sorted in descending order are
$p_1 = \frac{3}{4}$ and $p_2 = \frac{1}{4}$. Following the original two-stage method,
consider the intermediate target density matrix 
$\widetilde{\rho}_{\rm target} = 
\frac{1}{4} |0\rangle \langle 0| + \frac{3}{4} |1\rangle \langle 1| =
\begin{pmatrix}
	3/4 & 0\\
	0 & 1/4
\end{pmatrix}$,
the corresponding intermediate target state 
$\widetilde{x}_{\rm target} = (0, 0, \frac{1}{2})$, and the constant incoherent control 
$\overline{n}(t) \equiv \frac{p_2}{p_1 - p_2} = \frac{1}{2}$.
The corresponding solution $\overline{x}$ of the system~(\ref{QS_Bloch_parametrization}) is found from (\ref{G1_when_control_n_is_p})--(\ref{G3_when_control_n_is_p}):
$\overline{x}_1(t) = e^{-\gamma t} \cos(\omega t)$, $\overline{x}_2(t) = -e^{-\gamma t} \sin(\omega t)$,
$\overline{x}_3(t) = \frac{1}{2} \left( 1 - e^{-2 \gamma t} \right)$, $t \in [0, \widehat{t}]$. We take into account that the GKSL master equation in the sense of LDL is an approximate
for describing relevant physical phenomena. That is why, when we consider the equation
\begin{eqnarray}
\| \overline{x}(\widehat{t}) - \widetilde{x}_{\rm target} \|_2 = 
\left(\sum\limits_{j=1}^3 (\overline{x}_j(\widehat{t}) - \widetilde{x}_{{\rm target},j})^2 \right)^{1/2}
= \varepsilon_{\rm 1stage}
\label{condition_for_computing_widehat_t_in_unmodified_first_stage}
\end{eqnarray}
for obtaining $\widehat{t}$, we restrict 
the consideration to $\varepsilon_{\rm 1stage} \in \{10^{-2}, 10^{-3} \}$.
For numerical solving this equation, the function NSolve
in Wolfram Mathematica is used.  Thus, for $\varepsilon_{\rm 1stage} = 10^{-2}$ and 
$\varepsilon_{\rm 1stage} = 10^{-3}$ we obtain, correspondingly,  
$\widehat{t}(\varepsilon_{\rm 1stage} = 10^{-2}) \approx 2303$ 
and $\widehat{t}(\varepsilon_{\rm 1stage} = 10^{-3}) \approx 3454$.

Because when we consider piecewise constant incoherent controls, we expand 
the class of constant controls, we expect that $\widehat{t}$, which is sufficient
for satisfying the condition 
\begin{eqnarray}
J_1(n) = g_1({\bf a})=
\| x(\widehat{t}) - \widetilde{x}_{\rm target} \|_2^2 \leq 
\varepsilon_{\rm 1stage}^2, 
\label{condition_for_modified_first_stage}
\end{eqnarray}
can be decreased; here $x(\widehat{t})$ is obtained by solving the system with an 
admissible piecewise constant incoherent control and zero coherent 
control; $\varepsilon_{\rm 1stage} \in \{10^{-2}, 10^{-3} \}$. 
For minimizing the objective function $g_1({\bf a})$ for a given~$\widehat{t}$, which is less than the values of $\widehat{t}$ obtained above via the unmodified method, 
we use GPM-2 in the form (\ref{initial_vector_a})---(\ref{GPM_projection}). The method was implemented in Python~3 language.

We do not pretend to minimize $\widehat{t}$ among such 
values of $\widehat{t}$ that are sufficient for providing 
$J_1(n) = g_1({\bf a}) \leq \varepsilon_{\rm 1stage}^2 = 10^{-4}$ or 
$J_1(n) = g_1({\bf a}) \leq \varepsilon_{\rm 1stage}^2 = 10^{-6}$
with GPM-2. We fix $\widehat{t} = 450$ that is significantly less than
$\widehat{t}(\varepsilon_{\rm 1stage} = 10^{-2}) \approx 2303$ 
and $\widehat{t}(\varepsilon_{\rm 1stage} = 10^{-3}) \approx 3454$.
For considering piecewise constant controls, we divide $[0, 450]$ into $N = 225$ parts.
For visualization of the resulting trajectory, we compute the functions
$x_j$, $j=1,2,3$ in a larger number of time instants. In this example, we consider the initial vector ${\bf a}^{(0)} = (0, 0, \dots, 0)$ that gives $g_1({\bf a}^{(0)}) \approx  0.42$. 

\begin{figure}[ht!]
	\centering
	\includegraphics[width=\linewidth]{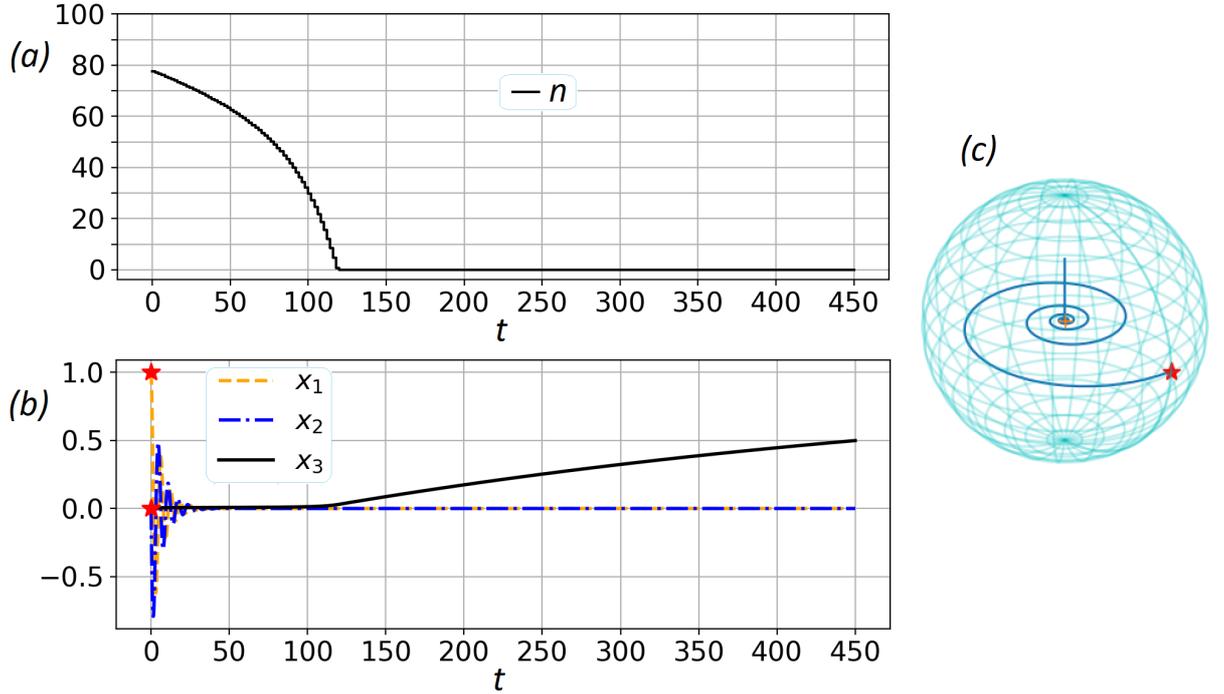} 
	\caption{For Example~1. Realization of the modified first stage of the two-stage method: steering from $x_0 = (1, 0, 0)$ to the point  close to the intermediate target state $\widetilde{x}_{\rm target} = (0, 0, 0.5)$ for $\widehat{t} = 450$ using GPM-2. The figures show the numerically optimized process~$(x,n)$: (a)~incoherent control~$n$ for $t \in [0, 450]$; (b)~functions $x_j$ of $t \in [0, 450]$, $j=1,2,3$; (c)~trajectory~$x$ in the Bloch ball. Here the distance $\| x(\widehat{t}) - \widetilde{x}_{\rm target} \|_2 \approx 10^{-3}$.} 
	\label{Figure2}
\end{figure}

For $\widehat{t} = 450$ and $\varepsilon_{\rm 1stage} = 10^{-3}$, 
GPM-2, which uses $\beta^{(m)} = 10$ $\forall~m \geq 0$ and $\lambda = 0.999$, provides the condition 
$J_1(n) = g_1({\bf a}) \leq \varepsilon_{\rm 1stage}^2 = 10^{-6}$
after 264 iterations by reaching $g_1({\bf a}) \approx 9.9 \cdot 10^{-7} \approx 10^{-6}$. Thus,
the distance $\| x(\widehat{t}) - \widetilde{x}_{\rm target} \|_2 \approx 10^{-3}$. Figure~\ref{Figure2} visualizes the results. For comparing, we use GPM-1 (i.e. always $\lambda = 0$) and after 1000 iterations we see that $g_1 \approx 0.0269$ that is significantly worse than the result of GPM-2 for the same algorithmic parameters expecting taking~$\lambda = 0$.

Further, for $\widehat{t} = 450$ and $\varepsilon_{\rm 1stage} = 10^{-2}$, 
we consider the iterations, which were carried out above with GPM-2 for
$\varepsilon_{\rm 1stage} = 10^{-3}$, and find that
the condition $J_1(n) = g_1({\bf a}) \leq \varepsilon_{\rm 1stage}^2 = 10^{-4}$
is satisfied after 82 iterations by reaching $g_1({\bf a}) \approx 9.7 \cdot 10^{-5} \approx 10^{-4}$. 

Moreover, consider $\widehat{t} = 400$, $N = 200$ for $\varepsilon_{\rm 1stage} = 10^{-2}$. Then GPM-2 with $\beta^{(m)} = 10$ and $\lambda = 0.999$ reaches $g_1({\bf a}) \approx 10^{-4}$
after 120 iterations. 

Comparing, on one hand, the results obtained via the modified first stage using GPM-2 and,
from other hand, the results of the unmodified first stage, we see that the modified first
stage can reach the same accuracy $\varepsilon_{\rm 1stage}$ significantly faster than the
unmodified first stage. If $\varepsilon_{\rm 1stage} = 10^{-3}$ and GPM-2 is used
when $\widehat{t} = 450$, then the difference of the durations is
$\widehat{t}(\varepsilon_{\rm 1stage} = 10^{-3}) / 450 \approx 3454 / 450 \approx 7.7$ times. 
If $\varepsilon_{\rm 1stage} = 10^{-2}$ and GPM-2 is used
when $\widehat{t} = 450$, then we find
$\widehat{t}(\varepsilon_{\rm 1stage} = 10^{-2}) / 450 \approx 2303 / 450 \approx 5.1$ times. 
If $\varepsilon_{\rm 1stage} = 10^{-2}$ and GPM-2 is used
when $\widehat{t} = 400$, then we find
$\widehat{t}(\varepsilon_{\rm 1stage} = 10^{-2}) / 450 \approx 2303 / 400 \approx 5.8$ times. 
However, this acceleration is achieved at the cost of considering the class of piecewise 
constant controls and the GPM-2 iterations. 

{\bf Example 2} (related to the example from the article~\cite[Sec.~V]{PechenPRA2011}). 
Consider the
initial state $x_0 = (0, 0, -1)$ (the Bloch ball's south pole) 
and the target state $x_{\rm target} = (0, 0, \frac{1}{2})$ 
corresponding to the target density matrix
$\rho_{\rm target} = \begin{pmatrix}
3/4 & 0 \\
0 & 1/4
\end{pmatrix}$. 
The intermediate target density matrix is
$\widetilde{\rho}_{\rm target} = 
\frac{3}{4} |0\rangle \langle 0| + \frac{1}{4} |1\rangle \langle 1| =
\begin{pmatrix}
	1/4 & 0\\
	0 & 3/4
\end{pmatrix}$ that corresponds to the state $\widetilde{x}_{\rm target} = (0, 0, -\frac{1}{2})$.
In the unmodified first stage of the method, we use the same constant control
$\overline{n} = \frac{1}{2}$ that is used in Example~1. Because $x_0$ is different from that in Example~1, 
here we find the corresponding solution $\overline{x}$ of the system~(\ref{QS_Bloch_parametrization}) from (\ref{G1_when_control_n_is_p})--(\ref{G3_when_control_n_is_p}):
$\overline{x}_1(t) = 0$, $\overline{x}_2(t) = 0$,
$\overline{x}_3(t) = \frac{1}{2} \left( 1 - 3 e^{-2 \gamma t} \right)$, $t \in [0, \widehat{t}]$. 
For this solution, consider the equation (\ref{condition_for_modified_first_stage}):
$\| \overline{x}(\widehat{t}) - \widetilde{x}_{\rm target} \|_2 = 
\left((\frac{1}{2} - \frac{3}{2} e^{-2\gamma \widehat{t}} + \frac{1}{2})^2 \right)^{1/2} = 
\left| 1 - \frac{3}{2} e^{-2\gamma \widehat{t}} \right|$.  
For $\varepsilon_{\rm 1stage} = 10^{-2}$ and $\varepsilon_{\rm 1stage} = 10^{-3}$, 
we obtain, correspondingly, $\widehat{t}(\varepsilon_{\rm 1stage} = 10^{-2}) \approx 98.9$
and $\widehat{t}(\varepsilon_{\rm 1stage} = 10^{-3}) \approx 101.1$ in the unmodified
first stage.

\begin{figure}[ht!]
\centering
\includegraphics[width=\linewidth]{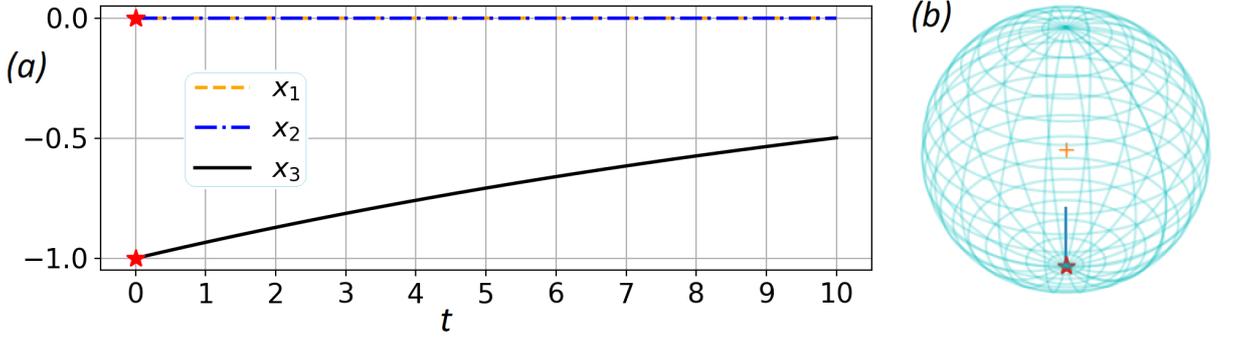} 
\caption{For Example~2. Realization of the modified first stage: (a)~functions $x_j$ of $t$; $j=1,2,3$; 
(b)~trajectory~$x$ in the Bloch~ball.} 
\label{Figure3}
\end{figure}

In the modified first stage, consider $\widehat{t} = 10$, $N = 1$ (i.e. without subintervals, see 
Corollary~\ref{corollary1}). We use GPM-2 with ${\bf a}^{(0)} = a^{(0)}_1 = 0$, 
$\beta^{(m)} = 1$, and $\lambda = 0.999$. The condition~(\ref{condition_for_modified_first_stage})
with $\varepsilon_{\rm 1stage} = 10^{-2}$ is satisfied via GPM-2 after 35 iterations.
The obtained control is $n \equiv 16.205$. This gives $J_1 \approx 4 \cdot 10^{-6} < \varepsilon_{\rm 1stage}^2 = 10^{-4}$.
Considering only $N=1$ significantly simplifies the computations, but at the same time this 
significantly decreases the first stage's duration in comparison to 
$\widehat{t}(\varepsilon_{\rm 1stage} = 10^{-2})$ of the unmodified first stage.
Figure~\ref{Figure3} shows the results. 

{\bf Example 3}. Consider the  
initial state $x_0 = (1, 0, 0)$ as in Example~1 and different target state $x_{\rm target} = (1/2, 0, 0)$ that corresponds to the target density matrix
$\rho_{\rm target} = \begin{pmatrix}
	1/2 & 1/4 \\
	1/4 & 1/2
\end{pmatrix}$ whose eigenvalues are $p_1 = 3/4$ and $p_2 = 1/4$. The intermediate target density matrix $\rho_{\rm target} = \begin{pmatrix}
	3/4 & 0 \\
	0 & 1/4
\end{pmatrix}$, which is considered in Example~1, has the same eigenvalues. In the current example, this matrix $\rho_{\rm target}$ and the results, which were obtained in Example~1 for the (un)modified first stage, are suitable.

\subsection{Realization of the second stage of the two-stage method}

{\bf Example 4} (in continuation of Example 1). The second stage of the method is implemented in the range $[\widehat{t}, T]$. 
At the moment $t = \widehat{t}$, we consider as the initial state $x(\widehat{t})$ 
the intermediate final state $x(\widehat{t}=450)$ computed at the modified first stage using the class of controls~(\ref{piecewise_constant_incoherent_control_1st_stage_open_system}), GPM-2, and $\varepsilon_{\rm 1stage} = 10^{-3}$. This state differs only very 
little from the intermediate target state $\widetilde{x}_{\rm target} = (0, 0, 0.5)$. 
At the second stage, with zero incoherent control ($n = 0$), we look for such final moment~$T$ 
and coherent control~$v$ that allow to obtain the final state $x(T)$ 
close to the given target state $x_{\rm target} = (0, 0, -0.5)$ with some suitable accuracy.

\begin{figure}[ht!]
	\centering
	\includegraphics[width=\linewidth]{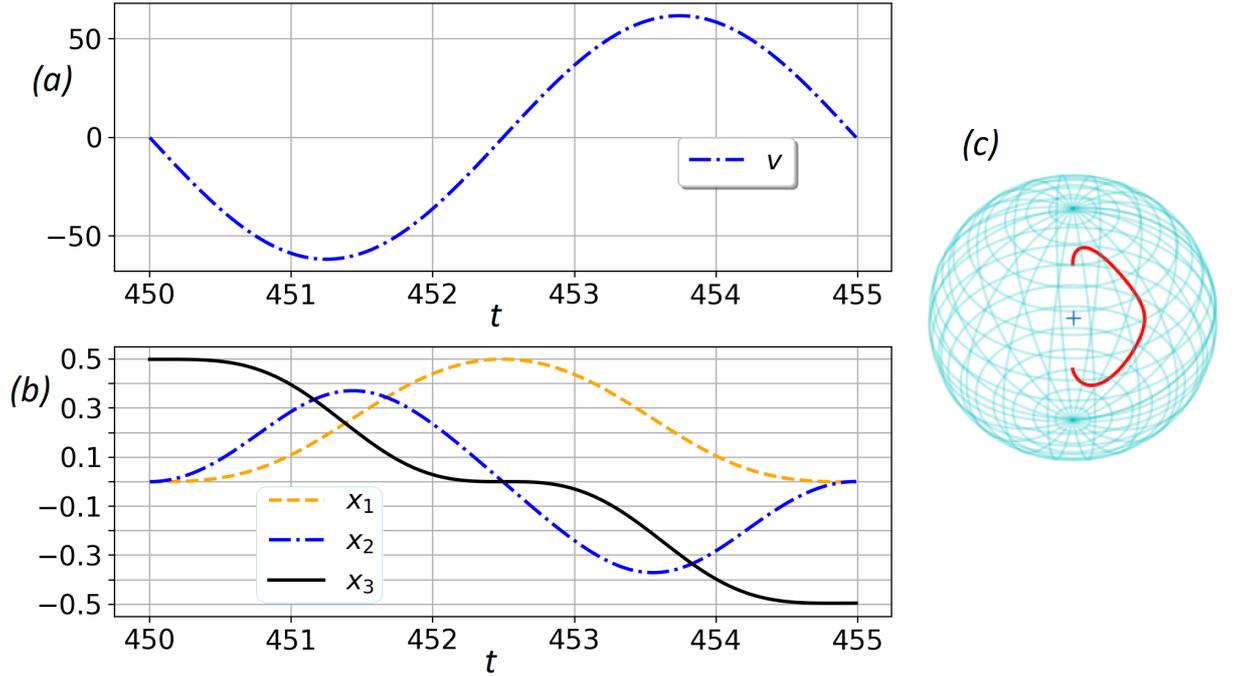} 
	\caption{For Example~4. In continuation of Example~1, realization of the second stage of the two-stage method: steering from the point, which is close to the intermediate target state $\widetilde{x}_{\rm target} = (0, 0, 0.5)$, to some point, which is close to the target state $x_{\rm target} = (0, 0, -0.5)$, during the time $T - \widehat{t} = 4.99$ under zero incoherent control and coherent control $v = -61.8 \sin\frac{2\pi (t - 450)}{4.99}$ of the type~(\ref{zero_ends_sin_coherent_control_2nd_stage_open_system}). Here are shown the resulting: (a)~control~$v$; (b)~functions $x_j$ of $t \in [450, 454.99]$; $j=1,2,3$; (c)~trajectory~$x$ in the Bloch~ball.} 
	\label{Figure4}
\end{figure}
The implementation of the second stage is performed in analogy with the example from the article \cite{PechenPRA2011}. Namely, we consider coherent control~$v$ of the type~(\ref{cos_coherent_control_2nd_stage_open_system}), at that we use the constraint $|A| \leq \nu = 100$ for the amplitude, and we set the frequency $\Omega = 1$.  Incoherent control is zero. The objective functional is $J_{2, {\rm trig.}}(v) = \| x(T) - x_{\rm target} \|_2 = \left( \sum\limits_{j=1}^3 (x_j(T) - x_{{\rm target},j})^2 \right)^{1/2} $. The final time~$T$ is defined as described below. On the range $[-\nu, \nu] = [-100, 100]$, we introduce a uniform grid  $\{A_j\}$ with the step $\Delta A = 0.05$. For each node $A_j$: 1)~we numerically solve the dynamical system with the controls $v = A_j \cos t$, $n = 0$; 2)~the corresponding vector-function~$x$ is stored on a uniform grid with nodes $\{t_k\}$ in the range $[\widehat{t}, \overline{T}]$ with step $\Delta t = 0.01$, where $\overline{T}$ is some sufficiently large number. After that, we look for such $\{t_k\}$ and $\{A_j\}$ that the distance $\| x(t_k) - x_{\rm target} \|_2 \leq \varepsilon_{\rm 2stage}$ with some small $0< \varepsilon_{\rm 2stage} \ll 1$, if, of course, such time nodes exist. We take $\varepsilon_{\rm 2stage} = 10^{-2}$. From the set of such nodes $\{t_k\}$, we select the node with the smallest value. We set $\overline{T} = 40$, i.e. the time grid is introduced in the range $[450, 490]$. By the described way, we obtain the amplitude $A = A_{j^{\ast}} = -67.6$ and the final time $T = t_{k^{\ast}} = 455.38$. The first stage's duration is $\widehat{t} = 450$, the second stage's duration is $455.38 - 450 = 5.38$. Here 
$\|x(T) - x_{\rm target}\|_2 \approx 4.5 \cdot 10^{-3} < 0.5\varepsilon_{\rm 2stage}$. 
Note that $v(\widehat{t}) \approx 49$ and $v(T) \approx 67$ here.

Further, consider the condition $v(\widehat{t}) = v(T) = 0$ 
and the class of controls~(\ref{zero_ends_sin_coherent_control_2nd_stage_open_system}).
With $\varepsilon_{\rm 2stage} = 10^{-2}$, we obtain $d = 2$, $T = 454.99$, and $A = -61.8$ that give
the distance $\|x(T) - x_{\rm target}\|_2 \approx 5 \cdot 10^{-3} \approx  0.5\varepsilon_{\rm 2stage}$.
Figure~\ref{Figure4} illustrates the resulting control process for the second stage.

Another version for realization of the second stage 
can be the following. With a given~$T$, consider the objective functional 
$J_2^{\alpha}(v) = \| x(T) - x_{\rm target} \|^2_2 + \alpha \int\limits_{\widehat{t}}^T v^2(t) / S(t)dt$ to be minimized, where
$v$ is piecewise continuous coherent control; $x$~is the solution of (\ref{QS_Bloch_parametrization}) with an admissible~$v$ and $n=0$; 
the integral term with $\alpha>0$, the function $S(t) = \exp\left(-b \left(\frac{t - \widehat{t}}{T - \widehat{t}} - \frac{1}{2} \right)^2 \right)$ with $b > 0$ is for approximate providing the condition $v(\widehat{t}) = v(T) = 0$ (this term is taken by analogy with such the possibility in the Krotov method's technique~\cite{MorzhinUMN2019}). 
With respect to numerical optimization of~$v$ via a gradient method, 
consider the Fr\'{e}chet derivative of $J_2^{\alpha}(v)$ using the general form for such derivative known 
in the theory of optimal control~\cite{DemyanovBook1970, VasilievBook2011}. 
The increment formula for $J_2^{\alpha}(v)$ on a given $v^{(m)}$ ($m \geq 0$) and arbitrary~$v$ is 
$J_2^{\alpha}(v) - J_2^{\alpha}(v^{(m)}) = \int\limits_{\widehat{t}}^T \left\langle \frac{\delta J_2^{\alpha}(v)}{\delta v(t)}\Big|_{v=v^{(m)}}, v(t) - v^{(m)}(t) \right\rangle dt +~r$, where $\frac{\delta J_2^{\alpha}(v)}{\delta v(t)}\Big|_{v=v^{(m)}} = -2 \mu \left(p_3^{(m)}(t) x_2^{(m)}(t) - p_2^{(m)}(t) x_3^{(m)}(t) \right) + 2\alpha \frac{v^{(m)}(t)}{S(t)}$; $r$ is the remainder; the functions $x^{(m)}$ 
and $p^{(m)}$ are the solutions, correspondingly, of (\ref{QS_Bloch_parametrization}), where the initial state 
$x(\widehat{t})$, controls $v = v^{(m)}$, $n = 0$ are used, and of the conjugate system 
$\frac{dp^{(m)}(t)}{dt} = -\left(A^{\rm T} + (B^v)^{\rm T} v^{(m)}(t) \right) p^{(m)}(t)$,  $p^{(m)}(T) = -2(x^{(m)}(T) - x_{\rm target})$. 

\section{Conclusions}
\label{section5_Conclusion}

The article considers control problems for some closed and open two-level quantum systems. 
The dynamics of the closed system is determined by the Schr\"{o}dinger equation with 
coherent control, while the dynamics of the open system is determined by the GKSL 
master equation whose Hamiltonian depends on coherent control and superoperator 
of dissipation depends on incoherent control.

For the closed system, 
the problem of generating the phase shift quantum gate for some set of phases and final times 
is studied. It is shown numerically (via the GRAPE-type approach and the approach using the dual annealing and differential evolution methods) that zero coherent control, which is a stationary point of the objective functional, is not optimal that gives an example of subtle point for practice in solving quantum control problems. Search based on exact analytical expression for the gradient is a suitable choice for revealing fine details of the quantum control landscape. 

For the open system, a modified version of the two-stage method which was proposed and developed for approximate generation of a target density matrix for generic $N$-level quantum systems in~\cite{PechenPRA2011}, in this work is studied for two-level systems with the goal of minimizing time necessary for the first (incoherent) stage: at the first stage, instead of constant incoherent control, which is explicitly specified using eigenvalues of the target density matrix, numerical optimization of piecewise constant incoherent control is carried out. Exact analytical formulas are obtained for the quantum system's state at the end of the first stage, the objective functions and their gradients depending on the parameters of piecewise constant controls. These formulas are used in the two-step GPM. For some values of the system parameters, the initial and target density matrices, various accuracies for the distance, in the numerical experiments it is shown that the durations of the modified first stage can be significantly less (in several times) than of the unmodified first stage, but at the cost of optimizing in the class of piecewise constant controls and using time-dependent incoherent controls. If the simplicity and efficiency of obtaining the constant incoherent control is more important than decreasing the duration of the first stage, then the unmodified method should be used. The choice of the unmodified method can also be caused by the fact that the constant incoherent control is simpler than some piecewise constant control. Otherwise if there is no experimental difficulty in working with piecewise constant controls and with applying an appropriate optimization method (for example, GPM-2) and if it is necessary to decrease the duration of the first stage, then one can use the modified version. 

\section*{Acknowledgements}
This work was funded by Russian Federation represented by the Ministry of Science and Higher Education (grant number 075-15-2020-788).

\end{document}